\documentclass[reqno]{amsart}
\usepackage{color}
\usepackage[mathscr]{eucal}
\usepackage{accents}
\usepackage[margin=2.5cm]{geometry}
\usepackage{tikz}
\usepackage{graphicx}
\usepackage{caption}
\usepackage{subcaption}

\newtheorem{theorem}{Theorem}[section]

\newtheorem{proposition}[theorem]{Proposition}
\newtheorem{corollary}[theorem]{Corollary}

\newtheorem{definition}{Definition}[section]
\newtheorem{remark}[theorem]{Remark}

\newcommand{\ut}{\widetilde u}
\newcommand{\uh}{\widehat u}
\newcommand{\utt}{\widetilde{\widetilde u}}

\newcommand{\uth}{\widehat{\widetilde u}}
\newcommand{\uo}{\overline u}

\newcommand{\cUt}{\widetilde\cU}
\newcommand{\cUh}{\widehat\cU}

\newcommand{\cUo}{\overline{\cU}}

\newcommand{\wt}{\widetilde}
\newcommand{\wh}{\widehat}
\newcommand{\ol}{\overline}
\newcommand{\mr}{\mathring}

\newcommand{\vp}{\varphi}

\newcommand{\rvp}{\mathring{\varphi}}

\newcommand{\up}{\Upsilon}
\newcommand{\lam}{\Lambda}
\newcommand{\az}{\mathtt A}
\newcommand{\bz}{\mathtt B}
\newcommand{\pot}{\upsilon}

\newcommand{\rpsi}{\mathring{\psi}}

\newcommand{\pz}{\mathtt p}

\newcommand{\cz}{\mathtt c}
\newcommand{\dz}{\mathtt d}
\newcommand{\Pz}{\mathtt P}

\newcommand{\po}{\overline{\pz}}
\newcommand{\Po}{\overline{\Pz}}
\newcommand{\poto}{\overline\upsilon}

\newcommand{\bphi}{\boldsymbol \phi}

\newcommand{\rlam}{\mathring{\Lambda}}
\newcommand{\rup}{\mathring{\Upsilon}}

\newcommand{\lamo}{\overline{\Lambda}}
\newcommand{\lamoo}{\overline{\lamo}}

\newcommand{\rN}{\rm N}

\newcommand{\cJ}{\mathcal J}
\newcommand{\cI}{\mathcal I}
\newcommand{\cA}{\mathscr A}
\newcommand{\cB}{\mathscr B}
\newcommand{\cC}{\mathscr C}
\newcommand{\cD}{\mathscr D}
\newcommand{\cF}{\mathcal F}
\newcommand{\cR}{\mathcal R}
\newcommand{\cK}{\mathscr K}

\newcommand{\cU}{\mathscr U}

\newcommand{\cO}{\mathcal O}

\newcommand{\asf}{\accentset{\boldsymbol \frown}}
\newcommand{\fru}{\asf{u}}

\newcommand{\frU}{\asf{\cU}}

\newcommand{\xx}{\hspace{0.5mm}}

\DeclareMathAccent{\wtilde}{\mathord}{largesymbols}{"65}
\DeclareMathAccent{\what}{\mathord}{largesymbols}{"62}

\numberwithin{equation}{section}
\Large
\begin{document}
\title{A Discrete Inverse Scattering Transform for Q3$_\delta$}
\author{Samuel Butler}
\address{Department of Applied Mathematics, 526 UCB, University of Colorado, Boulder, CO 80309-0526}
\email{samuel.butler@colorado.edu}
\date{October 2012}

\begin{abstract}
We derive a fully discrete Inverse Scattering Transform as a method for solving the initial-value problem for the Q3$_\delta$ lattice (difference-difference) equation for real-valued solutions. The initial condition is given on an infinite staircase within an N-dimensional lattice and must obey a given summability condition. The forward scattering problem is one-dimensional and the solution to Q3$_\delta$ is expressed through the solution of a singular integral equation. The solutions obtained depend on N discrete independent variables and N parameters.
\end{abstract}

\maketitle	
\section{introduction}

The Q3$_\delta$ equation
\begin{equation}\label{eq:Q3}
P(u\xx\uh+\ut\;\uth\xx)-Q(u\xx\ut+\uh\;\uth\xx)-(p^2-q^2)\left((\xx\ut\;\uh+u\xx\uth\xx)+\frac{\delta^2}{4PQ}\right)=0,
\end{equation}
where $P^2=(p^2-a^2)(p^2-b^2)$ and $Q^2=(q^2-a^2)(q^2-b^2)$, is a nonlinear lattice (difference-difference) equation. The dependent variable $u$ depends on two discrete independent variables $n,m \in \mathbb{Z}$ in the following way
\[
u=u(n,m) \hspace{1cm} \ut=u(n+1,m) \hspace{1cm} \uh=u(n,m+1) \hspace{1cm} \uth=u(n+1,m+1).
\]
The lattice parameters $p,q$ are associated with the $n$- and $m$-directions respectively, and $a,b,\delta$ are additional parameters. In 2003 Equation \eqref{eq:Q3} appeared in the ABS classification list \cite{abs:03} of such scalar affine-linear partial difference equations defined on four points, which possess the multidimensional consistency property \cite{nw:01} \cite{bs:02}. The particular parametrisation given here is due to \cite{nah:09} who also gave $N$-soliton solutions to \eqref{eq:Q3} using a Cauchy matrix approach (see also \cite{na:10} \cite{hz:09}). The extra parameter $\delta$ may be set to 1 without loss of generality, and in the special case $\delta=0$ equation \eqref{eq:Q3} can be (gauge) transformed to the NQC equation \cite{nqc:83}, which is a particular discretisation of a degenerate form of the Krichever-Novikov equation \cite{kn:80}. 
\newline
\par
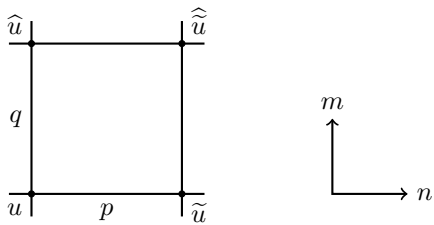
\begin{figure}[h!]
\begin{tikzpicture}[scale=1]
\draw [thick] (-0.3,0) -- (2.3,0);
\draw [thick] (-0.3,2) -- (2.3,2);
\draw [thick] (0,-0.3) -- (0,2.3);
\draw [thick] (2,-0.3) -- (2,2.3);
\draw [fill] (0,0) circle [radius=0.04];
\draw [fill] (0,2) circle [radius=0.04];
\draw [fill] (2,0) circle [radius=0.04];
\draw [fill] (2,2) circle [radius=0.04];
\node [below left] at (0,0) {$u$};
\node [above right] at (2,2) {$\uth$};
\node [below right] at (2,0) {$\ut$};
\node [above left] at (0,2) {$\uh$};
\node [below] at (1,0) {$p$};
\node [left] at (0,1) {$q$};
\draw [thick, <->] (4,1) -- (4,0) -- (5,0);
\node [above] at (4,1) {$m$};
\node [right] at (5,0) {$n$};
\end{tikzpicture}
\caption{Elementary lattice quadrilateral}
\label{fig:square}
\end{figure}

Equation \eqref{eq:Q3} defines a nonlinear relation between four points of the lattice shown in Figure \ref{fig:square}, and given its affine-linearity, may be solved uniquely for each point in terms of the other three.

This property allows one to define a well-posed initial value problem \cite{av:04} \cite{pnc:90} \cite{qcpn:91} by giving an initial profile along an infinite staircase within the $(n,m)$-lattice (see Figure \ref{fig:staircase1}). Due to the multidimensional consistency of \eqref{eq:Q3} however, this concept can be generalised. If we consider Q3$_\delta$ living in a multidimensional lattice, with copies of the equation imposed on each elementary quadrilateral within this space, then we can still set up a well-defined initial value problem by giving an initial profile along some {\bf multidimensional staircase $\Gamma$}. This is shown in Figure \ref{fig:staircase2}.

\begin{figure}[h!]
\begin{minipage}{0.3\textwidth}
\begin{tikzpicture}[scale=1]
\draw  (-0.3,0) -- (4.3,0);
\draw  (-0.3,4) -- (4.3,4);
\draw  (0,-0.3) -- (0,4.3);
\draw  (4,-0.3) -- (4,4.3);
\draw  (-0.3,1) -- (4.3,1);
\draw  (-0.3,3) -- (4.3,3);
\draw  (1,-0.3) -- (1,4.3);
\draw  (3,-0.3) -- (3,4.3);
\draw  (2,-0.3) -- (2,4.3);
\draw  (-0.3,2) -- (4.3,2);
\draw [ultra thick] (-0.3,0) -- (0,0) -- (0,1) -- (1,1) -- (1,2) -- (2,2) -- (2,3) -- (3,3) -- (3,4) -- (4,4) -- (4,4.3);
\end{tikzpicture}
\subcaption{Staircase of initial values in the $(n,m)$-lattice}
\label{fig:staircase1}
\end{minipage}
\hspace{1cm}
\begin{minipage}{0.3\textwidth}
\begin{tikzpicture}[scale=1]
\draw  (0,0) -- (0.7,0.7) -- (0.7,1.7) -- (0,2.4) -- (-0.7,1.7) -- (-0.7,0.7) -- (0,0) -- (0,1) -- (0.7,1.7) -- (0,1) -- (-0.7,1.7);
\draw  (1.4,0) -- (2.1,0.7) -- (2.1,1.7) -- (1.4,2.4) -- (0.7,1.7) -- (0.7,0.7) -- (1.4,0) -- (1.4,1) -- (2.1,1.7) -- (1.4,1) -- (0.7,1.7);
\draw  (2.8,0) -- (3.5,0.7) -- (3.5,1.7) -- (2.8,2.4) -- (2.1,1.7) -- (2.1,0.7) -- (2.8,0) -- (2.8,1) -- (3.5,1.7) -- (2.8,1) -- (2.1,1.7);
\draw  (-0.7,-1.7) -- (0,-1) -- (0,0) -- (-0.7,0.7) -- (-1.4,0) -- (-1.4,-1) -- (-0.7,-1.7) -- (-0.7,-0.7) -- (0,0) -- (-0.7,-0.7) -- (-1.4,0);
\draw  (0.7,-1.7) -- (1.4,-1) -- (1.4,0) -- (0.7,0.7) -- (0,0) -- (0,-1) -- (0.7,-1.7) -- (0.7,-0.7) -- (1.4,0) -- (0.7,-0.7) -- (0,0);
\draw  (2.1,-1.7) -- (2.8,-1) -- (2.8,0) -- (2.1,0.7) -- (1.4,0) -- (1.4,-1) -- (2.1,-1.7) -- (2.1,-0.7) -- (2.8,0) -- (2.1,-0.7) -- (1.4,0);
\draw  (3.5,-1.7) -- (4.2,-1) -- (4.2,0) -- (3.5,0.7) -- (2.8,0) -- (2.8,-1) -- (3.5,-1.7) -- (3.5,-0.7) -- (4.2,0) -- (3.5,-0.7) -- (2.8,0);
\draw [ultra thick] (-0.7,1.7) -- (0,1) -- (0,0) -- (0.7,0.7) -- (1.4,0) -- (2.1,0.7) -- (2.8,0) -- (2.8,-1) -- (3.5,-1.7) -- (4.2,-1);
\end{tikzpicture}
\subcaption{Multidimensional staircase $\Gamma$ of initial values}
\label{fig:staircase2}
\end{minipage}
\caption{}
\end{figure}

In this paper we solve the inital value problem for \eqref{eq:Q3} for an initial condition given along a multidimensional staircase $\Gamma$, within an N-dimensional lattice.

\subsection{History}$\\$

The Inverse Scattering Transform (IST) has been widely used as a mathematical tool for obtaining solutions of integrable nonlinear partial differential equations since its discovery in the 1960s by Gardner, Greene, Kruskal and Miura \cite{ggkm:67} \cite{ggkm:74}. Here the authors used it to solve the initial-value problem for the KdV equation, and the method has since seen many generalisations. Some other physically relevant partial differential equations which are also solvable by the IST include the nonlinear Schr\"odinger equation \cite{zs:72} and the sine-Gordon equation \cite{akns:73}. 
\par
The first application of the IST to differential-difference equations dates back to 1973 with Case and Kac \cite{ck:73} \cite{c:73}, who considered a discretisation of the Schr\"odiner equation on the half-line $n>0$. In 1974 Flaschka \cite{f:74} showed how this could be applied to solutions of the Toda lattice and Ablowitz and Ladik \cite{al:75} \cite{al:76} \cite{al:77} then derived a new dicrete scattering problem and showed how it was applicable to a number of physically important systems.
\par
More recently in 1999 Shabat \cite{s:99} found a new discrete scattering problem by applying Darboux transformations to the Schr\"odinger equation. This problem has since been studied by various authors \cite{bpps:01} \cite{r:02} \cite{lp:07} for the cases of continuous and discrete time evolution, and as an eigenvalue problem for analytic difference operators. This scattering problem also appears in the rigorous formulation of the discrete IST for the lattice potential KdV equation given in \cite{bj:10}, and was also generalised to a multidimensional setting in \cite{b:12}. While \cite{b:12} does consider the IST for Q3$_\delta$, there it was required that the ``potential" term arising from the initial condition decay exponentially. Here however the potential term arising from the initial condition must only decay faster than $n^{-2}$, which greatly widens the class of solutions obtained.

\subsection{Outline of Results}$\\$

In this paper we solve the initial value problem for the Q3$_\delta$ equation \eqref{eq:Q3} by developing a fully discrete IST for this equation which naturally incorporates its multidimensional consistency. We begin in Section \ref{sec:linearproblem} by giving the Lax pair for \eqref{eq:Q3} and looking at the associated linear problem arising from this. In Section \ref{sec:motivation} we then set up the IST framework (reality assumptions, initial value space, boundary conditions, etc.) within the multidimensional lattice. The forward scattering problem is carried out in Section \ref{sec:phi1forward}, where Jost solutions are constructed as functions existing on the staircase $\Gamma$. An important step is then given in Section \ref{sec:timeevo}, where we make assumptions on the boundary conditions for the solutions off $\Gamma$. This is essentially imposing that our boundary conditions are ``time" independent. The inverse problem is then carried out in Sections \ref{sec:phi1inverse} and \ref{sec:phi2inverse}. We then show in Section \ref{sec:ureconstruction} how to reconstruct the solution $u$ of Q3$_\delta$. A key part of this procedure is the solving of the singular integral equation
\begin{equation}
\xi_{\pm b}(\zeta)=\frac1{\zeta\pm b}-\sum_{k=1}^M\left(\frac{\cz_k\xx\xi_{\pm b}(\zeta_k)}{\zeta+\zeta_k}\right)\rho(\zeta_k)-\frac1{2\pi i}\int_{-i\infty}^{+i\infty}\left(\frac{R(\sigma)\xx\xi_{\pm b}(\sigma)}{\sigma+\zeta}\right)\rho(\sigma)\xx d\sigma.
\end{equation}
for the function $\xi_{\pm b}(\zeta)$, where the plane-wave factors are contained in $\rho$ and all other quantities appearing in the equation are known in terms of the scattering data. Finally a one-soliton example is given in Section \ref{sec:onesoliton}. For convenience some of the longer proofs are given in the Appendix.

\section{Linear Problem for Q3$_\delta$}\label{sec:linearproblem}

A Lax pair for equation \eqref{eq:Q3}, which was first given in \cite{na:10}, is
\begin{subequations}\label{eq:laxnm}
\begin{align}
(p^2-\zeta^2)^{\frac12}\wt{\bphi}&=\frac1{\cU}\left(\begin{array}{cc} P\xx\ut-(p^2-b^2)u & \zeta^2-b^2 \vspace{0.1 in}\\ \cU\cUt-\frac{\delta^2(p^2-b^2)}{4P(\zeta^2-b^2)} & (p^2-b^2)\xx\ut-Pu \end{array}\right)\bphi \\
(q^2-\zeta^2)^{\frac12}\wh{\bphi}&=\frac1{\cU}\left(\begin{array}{cc} Q\xx\uh-(q^2-b^2)u & \zeta^2-b^2 \vspace{0.1 in}\\ \cU\cUh-\frac{\delta^2(q^2-b^2)}{4Q(\zeta^2-b^2)} & (q^2-b^2)\xx\uh-Qu \end{array}\right)\bphi,
\end{align}
\end{subequations}
where this system is consistent (i.e. $\wh{\wt\bphi}=\wt{\wh\bphi}$) if and only if $u$ solves \eqref{eq:Q3}. Here $\zeta$ is the spectral parameter. The dual function $\cU$ is determined by solving the first-order equations
\begin{subequations}
\begin{align}
\cU\cUt&=P(u^2+\ut\xx^2)-(2p^2-a^2-b^2)u\xx\ut+\frac{\delta^2}{4P} \\
\cU\cUh&=Q(u^2+\uh\xx^2)-(2q^2-a^2-b^2)u\xx\uh+\frac{\delta^2}{4Q}.
\end{align}
\end{subequations}
These equations can be solved to determine $\cU$. The integration constants are determined by the boundary conditions that will be imposed on $\cU$. This is the setup for the linear problem of Q3$_\delta$ in the $(n,m)$-plane. In order to incorporate the multidimensional consistency of this equation however, we consider how the Lax equations \eqref{eq:laxnm} change when we we make a shift in any one of the N possible directions within the lattice. Let the N discrete independent variables be denoted by $n_1, ..., n_{\rN}$ (all elements of $\mathbb{Z}$), with associated parameters $p_1, ..., p_{\rN}$. If we choose an arbitrary direction with independent variable $n_k$ and associated parameter $p_k$, then we let $\fru$ denote a shift of $u$ in this $n_k$-direction, that is
\begin{align*}
u&=u(n_1,n_2,...,n_k,...,n_N;p_1,p_2,...,p_k,...p_{\rN}) \\
\fru&=u(n_1,n_2,...,n_k+1,...,n_N;p_1,p_2,...,p_k,...p_{\rN}).
\end{align*}
Due to the symmetry of the equation and its multidimensional consistency, the Lax equation is this $n_k$-direction is
\begin{equation}\label{eq:Laxk}
(p_k^2-\zeta^2)^{\frac12}\asf{\bphi}=\frac1{\cU}\left(\begin{array}{cc} P_k\fru-(p_k^2-b^2)u & \zeta^2-b^2 \vspace{0.1 in}\\ \cU\frU-\frac{\delta^2(p_k^2-b^2)}{4P_k(\zeta^2-b^2)} & (p_k^2-b^2)\fru-P_ku \end{array}\right)\bphi.
\end{equation}
Equation \eqref{eq:Laxk} is then compatible with every other of the N-1 remaining Lax equations (we have $\frac12$N(N-1) Lax pairs), provided that $u$ solves Q3$_\delta$ in each pair of lattice directions. The dependence of $\cU$ on $n_k$ is found by solving
\begin{equation}\label{eq:cU}
\cU\frU=P_k(u^2+\fru^2)-(2p_k^2-a^2-b^2)u\fru+\frac{\delta^2}{4P_k},
\end{equation}
and again using the boundary conditions of $\cU$ to determine the integration constants.

\section{Setting up the Discrete IST}\label{sec:motivation}

Here we set up the discrete IST in the multidimensional lattice.

\subsection{Reality Assumptions} $\\$

While equation \eqref{eq:Q3} is defined for complex-valued solutions $u$, for the purposes of the IST we assume that
\begin{enumerate}
\item[-] The solution $u$ and all parameters $a, b, p_1, ..., p_{\rN}$ are {\bf real}.
\item[-] Since the equation only depends on the squares $a^2, b^2, p_1^2, ..., p_{\rN}^2$, we choose to set $a>b>0$ and $p_k>0$ for all $k=1, ..., \rN$.
\end{enumerate}
Complex-valued solutions were allowed in \cite{b:12}, however this led to a significantly stronger restriction on the initial condition. Our aim here is to obtain the widest possible class of solutions.

\subsection{Staircase} $\\$

We now define the staircase $\Gamma$ of initial conditions. Let us first consider some examples. 
\par
Suppose first that we choose to give the initial condition on a line $\Gamma_o$ spanned by the variable $n$ (and parameter $p$), which corresponds to one of the N variables $n_k$ (with parameter $p_k$). This is shown in Figure \ref{fig:staircase3}.
\begin{figure}[h!]
\begin{minipage}{0.3\textwidth}
\begin{tikzpicture}[scale=1]
\draw  (-0.3,0) -- (4.3,0);
\draw  (-0.3,4) -- (4.3,4);
\draw  (0,-0.3) -- (0,4.3);
\draw  (4,-0.3) -- (4,4.3);
\draw  (-0.3,1) -- (4.3,1);
\draw  (-0.3,3) -- (4.3,3);
\draw  (1,-0.3) -- (1,4.3);
\draw  (3,-0.3) -- (3,4.3);
\draw  (2,-0.3) -- (2,4.3);
\draw  (-0.3,2) -- (4.3,2);
\draw [ultra thick,<->] (-0.3,0) -- (4.3,0);
\node [above right] at (2,0) {$\Gamma_o$};
\end{tikzpicture}
\subcaption{Line $\Gamma_o$ of initial values}
\label{fig:staircase3}
\end{minipage}
\hspace{1cm}
\begin{minipage}{0.3\textwidth}
\begin{tikzpicture}[scale=1]
\draw  (-0.3,0) -- (4.3,0);
\draw  (-0.3,4) -- (4.3,4);
\draw  (0,-0.3) -- (0,4.3);
\draw  (4,-0.3) -- (4,4.3);
\draw  (-0.3,1) -- (4.3,1);
\draw  (-0.3,3) -- (4.3,3);
\draw  (1,-0.3) -- (1,4.3);
\draw  (3,-0.3) -- (3,4.3);
\draw  (2,-0.3) -- (2,4.3);
\draw  (-0.3,2) -- (4.3,2);
\draw [ultra thick,<->] (-0.3,0) -- (0,0) -- (0,1) -- (1,1) -- (1,2) -- (2,2) -- (2,3) -- (3,3) -- (3,4) -- (4,4) -- (4,4.3);
\node [above right] at (2,2) {$\Gamma_1$};
\end{tikzpicture}
\subcaption{$(1,1)$ staircase $\Gamma_1$ of initial values}
\label{fig:staircase4}
\end{minipage}
\caption{}
\end{figure}
This is the usual setup for the continuous IST for equations such as the KdV equation. Along this line all other N-1 variables are held constant. From the Lax equation \eqref{eq:Laxk}, by eliminating the second component, the first component of $\bphi$ satisfies
\[
(p^2-\zeta^2)^{\frac12}\xx\wt{\wt\phi}_1-\left(\frac{P(\xx\utt-u)}{\cUt}\right)\wt{\phi}_1+(p^2-\zeta^2)^{\frac12}\xx\phi_1=0.
\]
The second component in then constructed from
\[
(\zeta^2-b^2)\xx\phi_2=(p^2-\zeta^2)^{\frac12}\xx\cU\xx\wt\phi_1-(P\xx\ut-(p^2-b^2)u)\xx\phi_1.
\]
This gives $\bphi$ as a function of $n$ along the line $\Gamma_o$, which is the direct scattering problem. In order to obtain $\bphi$ as a function of the remaining N-1 lattice variables, we need to give boundary conditions at $n\to-\infty$ or $n\to+\infty$, and then use the remaining N-1 Lax equations \eqref{eq:Laxk} to determine the ``time" evolution of the scattering data in each of these directions. In the language of the continuous theory this discrete IST is a 1+(N-1)-type scattering problem. 
\par
Contrary to the continuous case, one of the benefits of the discrete setup is that we can easily change the one-dimensional manifold along which we specify the initial conditions. Perhaps the most natural such manifold is a (1,1)-staircase in the $(n,m)$-plane, shown in Figure \ref{fig:staircase4}. For each 3-point segment of the staircase which iterates first in the $n$-direction and then in the $m$-direction we have
\[
(q^2-\zeta^2)^{\frac12}\xx\wh{\wt\phi}_1-\left(\frac{Q\xx\uth-(q^2-p^2)\xx\ut-Pu}{\cUt}\right)\wt{\phi}_1+(p^2-\zeta^2)^{\frac12}\xx\phi_1=0,
\]
while for each 3-point segment which iterates first in the $m$-direction and then in the $n$-direction
\[
(p^2-\zeta^2)^{\frac12}\xx\wh{\wt\phi}_1-\left(\frac{P\xx\uth-(p^2-q^2)\xx\uh-Qu}{\cUh}\right)\wh{\phi}_1+(q^2-\zeta^2)^{\frac12}\xx\phi_1=0.
\]
Solving these equations gives $\phi_1$, and then $\phi_2$ may be constructed from similarly considering the first component of the two Lax equations. Thus the result of the forward scattering problem is that we know $\bphi$ as a function along $\Gamma_1$, that is in terms of some independent staircase variable which depends on $n$ and $m$. If we let this new variable be denoted by $i$, and let $i_o$ correspond to the point $(n_o,m_o)$ on $\Gamma_1$, then assuming that we iterate first in the $n$-direction, the change of variables from $i$ to $n,m$ is given by
\begin{equation}\label{eq:cov}
n-n_o=\left\lfloor\xx\frac12(i+1-i_o)\xx\right\rfloor, \;\;\;\;\; m-m_o=\left\lfloor\xx\frac12(i-i_o)\xx\right\rfloor,
\end{equation}
where the brackets denote the floor function. This is once again a 1+(N-1)-type scattering problem.
\par
We now look at how this may be generalised. Consider an arbitrary staircase $\Gamma$ which has the following properties:
\begin{enumerate}
\item[-] $\Gamma$ is an infinite staircase which iterates in $\cI$ of the N lattice directions, where $1\leq\cI\leq\rN$
\item[-] Every iteration along $\Gamma$ corresponds to a positive iteration in one of the lattice variables $n_k$
\item[-] $\Gamma$ is defined through some repeated stepping algorithm.
\end{enumerate}
Figures \ref{fig:staircase5} and \ref{fig:staircase6} show examples of such staircases, where we assume that both of these are repeated infinitely in both directions. Figure \ref{fig:staircase7} shows an example of a staircase which violates the above criteria.
\begin{figure}[h!]
\begin{minipage}{0.25\textwidth}
\begin{tikzpicture}[scale=0.8]
\draw  (-0.3,0) -- (4.3,0);
\draw  (-0.3,4) -- (4.3,4);
\draw  (0,-0.3) -- (0,4.3);
\draw  (4,-0.3) -- (4,4.3);
\draw  (-0.3,1) -- (4.3,1);
\draw  (-0.3,3) -- (4.3,3);
\draw  (1,-0.3) -- (1,4.3);
\draw  (3,-0.3) -- (3,4.3);
\draw  (2,-0.3) -- (2,4.3);
\draw  (-0.3,2) -- (4.3,2);
\draw [ultra thick, <->] (0,-0.3) -- (0,0) -- (2,0) -- (2,2) -- (4,2) -- (4,4) -- (4.3,4);
\end{tikzpicture}
\subcaption{$(2,2)$ staircase of initial values}
\label{fig:staircase5}
\end{minipage}
\hspace{1cm}
\begin{minipage}{0.25\textwidth}
\begin{tikzpicture}[scale=0.8]
\draw  (0,0) -- (0.7,0.7) -- (0.7,1.7) -- (0,2.4) -- (-0.7,1.7) -- (-0.7,0.7) -- (0,0) -- (0,1) -- (0.7,1.7) -- (0,1) -- (-0.7,1.7);
\draw  (1.4,0) -- (2.1,0.7) -- (2.1,1.7) -- (1.4,2.4) -- (0.7,1.7) -- (0.7,0.7) -- (1.4,0) -- (1.4,1) -- (2.1,1.7) -- (1.4,1) -- (0.7,1.7);
\draw  (2.8,0) -- (3.5,0.7) -- (3.5,1.7) -- (2.8,2.4) -- (2.1,1.7) -- (2.1,0.7) -- (2.8,0) -- (2.8,1) -- (3.5,1.7) -- (2.8,1) -- (2.1,1.7);
\draw  (-0.7,-1.7) -- (0,-1) -- (0,0) -- (-0.7,0.7) -- (-1.4,0) -- (-1.4,-1) -- (-0.7,-1.7) -- (-0.7,-0.7) -- (0,0) -- (-0.7,-0.7) -- (-1.4,0);
\draw  (0.7,-1.7) -- (1.4,-1) -- (1.4,0) -- (0.7,0.7) -- (0,0) -- (0,-1) -- (0.7,-1.7) -- (0.7,-0.7) -- (1.4,0) -- (0.7,-0.7) -- (0,0);
\draw  (2.1,-1.7) -- (2.8,-1) -- (2.8,0) -- (2.1,0.7) -- (1.4,0) -- (1.4,-1) -- (2.1,-1.7) -- (2.1,-0.7) -- (2.8,0) -- (2.1,-0.7) -- (1.4,0);
\draw  (3.5,-1.7) -- (4.2,-1) -- (4.2,0) -- (3.5,0.7) -- (2.8,0) -- (2.8,-1) -- (3.5,-1.7) -- (3.5,-0.7) -- (4.2,0) -- (3.5,-0.7) -- (2.8,0);
\draw [ultra thick, <->] (-1.6,-0.8) -- (-1.4,-1) -- (-1.4,0) -- (-0.7,0.7) -- (0,0) -- (0.7,0.7) -- (1.4,0) -- (1.4,1) -- (2.1,1.7) -- (2.8,1) -- (3.5,1.7) -- (4.2,1) -- (4.2,1.4);
\end{tikzpicture}
\subcaption{Repeated three-dimensional staircase of initial values}
\label{fig:staircase6}
\end{minipage}
\hspace{1.5cm}
\begin{minipage}{0.25\textwidth}
\begin{tikzpicture}[scale=0.8]
\draw  (-0.3,0) -- (4.3,0);
\draw  (-0.3,4) -- (4.3,4);
\draw  (0,-0.3) -- (0,4.3);
\draw  (4,-0.3) -- (4,4.3);
\draw  (-0.3,1) -- (4.3,1);
\draw  (-0.3,3) -- (4.3,3);
\draw  (1,-0.3) -- (1,4.3);
\draw  (3,-0.3) -- (3,4.3);
\draw  (2,-0.3) -- (2,4.3);
\draw  (-0.3,2) -- (4.3,2);
\draw [ultra thick, <->] (0,-0.3) -- (0,0) -- (1,0) -- (1,1) -- (3,1) -- (3,3) -- (1,3) -- (1,4.3);
\end{tikzpicture}
\subcaption{Inadmissible staircase of initial values}
\label{fig:staircase7}
\end{minipage}
\caption{}
\end{figure}

\subsection{Linear Problem} $\\$

Given a staircase $\Gamma$ of initial values, we now look at the linear problem along $\Gamma$. Since each iteration along $\Gamma$ is in one of the lattice directions, the Lax equation for each iteration will be one of the $\cI$ equations \eqref{eq:Laxk}. From here it is convenient to introduce the staircase variable $i$, which will cycle through the various lattice variables encountered along $\Gamma$, and will be related to these $\cI$ lattice variables by a relation similar to \eqref{eq:cov}. To take care of the fact that the parameters $p_k$ will also change along the staircase we introduce the staircase parameter $\pz=\pz(i)$ (and $\Pz^2=(\pz^2-a^2)(\pz^2-b^2)$), which will cycle through the parameters $p_k$ encountered along the staircase. This new variable $i$ and parameter $\pz(i)$ allow the scattering problem for $\phi_1$ (obtained by eliminating $\phi_2$) along $\Gamma$ to be conveniently expressed as
\begin{equation}\label{eq:ds1}
(\po^2-\zeta^2)^{\frac12}\xx\ol{\ol\phi}_1-\left(\frac{\Po\xx\ol{\uo}-(\po^2-\pz^2)\xx\uo-\Pz\xx u}{\cUo}\right)\ol{\phi}_1+(\pz^2-\zeta^2)^{\frac12}\xx\phi_1=0,
\end{equation}
where
\begin{align*}
&\phi_1=\phi_1(i;\zeta), \;\;\;\;\; \ol{\phi}_1=\phi_1(i+1;\zeta), \;\;\;\;\; \pz=\pz(i), \;\;\;\;\; \po=\pz(i+1).
\end{align*}
The second component of the eigenfunction is then constructed from
\begin{equation}\label{eq:ds2}
(\zeta^2-b^2)\xx\phi_2=(\pz^2-\zeta^2)^{\frac12}\xx\cU\xx\ol\phi_1-(\Pz\xx\uo-(\pz^2-b^2)u)\xx\phi_1.
\end{equation}
The forward scattering problem, explored in the next Section, is the problem of solving equations \eqref{eq:ds1} and \eqref{eq:ds2} on $\Gamma$.

\subsection{Boundary Conditions} $\\$

The boundary conditions that we assume on the soluton $u$ are those exhibited by the known soliton solutions \cite{nah:09} of Q3$_\delta$. If we define
\begin{subequations}
\begin{align}
\rho(k):=&\;\prod_{r=1}^N\xx\left(\frac{p_r+k}{p_r-k}\right)^{n_r}, \;\;\;\;\; \cF(a,b):=\prod_{r=1}^N\left(\frac{(p_r+a)(p_r+b)}{(p_r-a)(p_r-b)}\right)^{\frac12n_r} \\
S(a,b):=&\;\frac{1+\left(\frac{(a-k)(b-k)}{(a+k)(b+k)}\right)\rho(k)}{1+\rho(k)}, \;\;\;\;\; V(a):=\;\frac{1+\left(\frac{a-k}{a+k}\right)\rho(k)}{1+\rho(k)}
\end{align}
\end{subequations}
then a one-soliton solution to Q3 (which depends on all N lattice variables) is given by
\begin{equation}
u=\cA\xx\cF(a,b)\xx S(a,b)+\cB\xx\cF(a,-b)\xx S(a,-b)+\cC\xx\cF(-a,b)\xx S(-a,b)+\cD\xx\cF(-a,-b)\xx S(-a,-b),
\end{equation}
where the four constants are restrained by
\begin{equation}\label{eq:abcd}
\cA\cD(a+b)^2-\cB\cC(a-b)^2=-\frac{\delta^2}{16ab}.
\end{equation}
The corresponding dual function $\cU$ is
\begin{align*}
\cU&=(a+b)\cA\xx\cF(a,b)\xx V(a)V(b)+(a-b)\cB\xx\cF(a,-b)\xx V(a)V(-b) \\
&-(a-b)\cC\xx\cF(-a,b)\xx V(-a)V(b)-(a+b)\cD\xx\cF(-a,-b)\xx V(-a)V(-b).
\end{align*}
Using this as a guide, and bearing in mind the choice $a>b>0$, the boundary conditions that we assume on the solution $u$ are
\begin{subequations}\label{eq:uBC}
\begin{align}\label{eq:uBC-infty}
&u\sim\cC\xx\cF(-a,b)+\cD\xx\cF(-a,-b) \;\;\;\; {\rm as} \; i\to-\infty \\
\label{eq:uBC+infty}
&u\sim\cK_o\cA\xx\cF(a,b)+\cK_1\cB\xx\cF(a,-b) \;\;\;\; {\rm as} \; i\to+\infty,
\end{align}
\end{subequations}
where $\cK_o$ and $\cK_1$ are constants and the plane-wave factors $\cF$ satisfy
\begin{equation}
\ol{\cF}(a,b)=\left(\frac{(\pz(i)+a)(\pz(i)+b)}{(\pz(i)-a)(\pz(i)-b)}\right)^{\frac12}\cF(a,b).
\end{equation}
The corresponding boundary conditions for $\cU$ are
\begin{subequations}\label{eq:cUBC}
\begin{align}\label{eq:cUBC-infty}
&\cU\sim-(a-b)\cC\xx\cF(-a,b)-(a+b)\cD\xx\cF(-a,-b) \;\;\;\; {\rm as} \; i\to-\infty \\
\label{eq:UBC+infty}
&\cU\sim\cK_o(a+b)\cA\xx\cF(a,b)+\cK_1(a-b)\cB\xx\cF(a,-b) \;\;\;\; {\rm as} \; i\to+\infty.
\end{align}
\end{subequations}

\section{Forward Scattering of $\phi_1$}\label{sec:phi1forward}

We are now in a position to carry out the forward scattering problem. We firstly consider the forward scattering problem for $\phi_1$, which is determining the solution of \eqref{eq:ds1}, given the above boundary conditions. With this choice of boundary conditions, by explicit calculation it follows that the function $\Omega$ defined by
\begin{equation}\label{eq:Omega}
\Omega:=\frac{\Po\xx\ol{\uo}-(\po^2-\pz^2)\xx\uo-\Pz\xx u}{\cUo}
\end{equation}
has the following asymptotic behaviour:
\begin{equation}
\Omega\sim\pz+\po \;\;\;\;\; {\rm as} \;\; i\to\pm\infty.
\end{equation}
This asymptotic result shows that the object $\Omega$ behaves like a difference of H1-type soliton solutions as $i\to\pm\infty$ (see e.g. \cite{bj:10}). This is perhaps not surprising as in \cite{nah:09} and \cite{na:10} Miura-type relations between soliton solutions of H1 and Q3$_\delta$	 were found, which take precisely the form of the quantity $\Omega$. This object is exactly the term appearing in \eqref{eq:ds1}, and we may therefore rewrite this equation as
\begin{equation}\label{eq:ds11}
(\po^2-\zeta^2)^{\frac12}\xx\ol{\ol\phi}_1-\bigl(\xx\pz+\po+\poto\xx\bigr)\xx\ol{\phi}_1+(\pz^2-\zeta^2)^{\frac12}\xx\phi_1=0.
\end{equation}
\begin{definition}\label{def:pot}
Given an initial condition $u=u(i)$ along the staircase $\Gamma$, the potential $\pot=\pot(i)$ is defined to be
\begin{align}\label{eq:pot}
\pot(i+1)\equiv\poto:=&\xx\Omega-\pz-\po=\left(\frac{\Po\xx\ol{\uo}-(\po^2-\pz^2)\xx\uo-\Pz\xx u}{\cUo}\right)-\pz-\po,
\end{align}
where $\cU$ is determined by \eqref{eq:cU}.
\end{definition}
With this definition the potential $\pot$ vanishes at either end of the staircase $\Gamma$. 

\begin{definition}\label{def:jost}
The Jost solutions $\vp,\rvp$ to \eqref{eq:ds11} are defined by the boundary conditions
\begin{subequations}
\begin{align}\label{eq:bcphis}
&\vp(i;\zeta)\sim\prod_{r=0}^{i-1}\left(\frac{\pz(r)+\zeta}{\pz(r)-\zeta}\right)^{\frac12} \;\; {\rm as} \;\; i\to-\infty \\
&\rvp(i;\zeta)\sim\prod_{r=0}^{i-1}\left(\frac{\pz(r)-\zeta}{\pz(r)+\zeta}\right)^{\frac12} \;\; {\rm as} \;\; i\to-\infty,
\end{align}
and the Jost solutions $\psi,\rpsi$ to equation \eqref{eq:ds11} are defined by the boundary conditions
\begin{align}\label{eq:bcpsis}
&\psi(i;\zeta)\sim\prod_{r=0}^{i-1}\left(\frac{\pz(r)-\zeta}{\pz(r)+\zeta}\right)^{\frac12} \;\; {\rm as} \;\; i\to+\infty \\
&\rpsi(i;\zeta)\sim\prod_{r=0}^{i-1}\left(\frac{\pz(r)+\zeta}{\pz(r)-\zeta}\right)^{\frac12} \;\; {\rm as} \;\: i\to+\infty.
\end{align}
\end{subequations}
\end{definition}

Since equation \eqref{eq:ds11} is invariant under the map $\zeta\to-\zeta$, it follows by the definition of the boundary conditions for the Jost solutions and uniqueness of the boundary value problem \cite{m:90}, that
\begin{equation}
\rvp(i;\zeta)=\vp(i;-\zeta), \;\;\; \rpsi(i;\zeta)=\psi(i;-\zeta).
\end{equation}
Since the general solution to \eqref{eq:ds11} involves two linearly independent solutions we may write
\begin{equation}\label{eq:ab}
\psi=\az\xx\rvp+\bz\xx\vp, \;\;\; \rpsi=\mr{\az}\xx\vp+\mr{\bz}\xx\rvp
\end{equation}
where $\az$ and $\bz$ are independent of $i$ and $\mr{\az}(\zeta)=\az(-\zeta)$ and $\mr{\bz}(\zeta)=\bz(-\zeta)$. 

\begin{proposition}\label{prop:aneq0}
If $\zeta$ is purely imaginary then
\begin{equation}\label{eq:aneq0}
|\az(\zeta)|^2=1+|\bz(\zeta)|^2.
\end{equation}
\begin{proof}
Firstly given any two solutions $x(i)$ and $y(i)$ of \eqref{eq:ds11}, by eliminating the potential term one can show that the Wronskian
\begin{equation}
W(x,y):=(\pz^2-\zeta^2)^{\frac12}(x\xx\ol{y}-\ol{x}\xx y)
\end{equation}
is independent of $i$. Furthermore if $\zeta$ is purely imaginary then equation \eqref{eq:ds11} is purely real and thus $\vp^*$ and $\psi^*$ (the complex conjugates of $\vp$ and $\psi$) are also solutions of this equation. By comparing the boundary conditions for $\vp^*,\rvp$ and $\psi^*,\rpsi$, by the uniqueness of the boundary value problem we have
\[
\vp^*(i;\zeta^*)\equiv\rvp(i;\zeta), \;\;\;\;\; \psi^*(i;\zeta^*)\equiv\rpsi(i;\zeta).
\]
By taking the complex conjugate of \eqref{eq:ab} we then have $\mr{\az}(\zeta)\equiv\az^*(\zeta^*)$ and $\mr{\bz}(\zeta)\equiv\bz^*(\zeta^*)$. Now due to the linearity and anti-symmetry of the Wronskian we have
\[
W(\psi,\psi^*)=W(\az\vp^*+\bz\vp,\az^*\vp+\bz^*\vp^*)=\bigl(|\az|^2-|\bz|^2\bigr)W(\vp^*,\vp),
\]
and since the Wronskian is independent of $i$, these may be evaluated at the relevant boundaries which gives $W(\psi,\psi^*)=W(\vp^*,\vp)=2\zeta$. This proves \eqref{eq:aneq0}.
\end{proof}
\end{proposition}

\subsection{Analyticity and Asymptoticity Properties of the Jost Solutions}

We now determine asymptoticity properties of the Jost solutions as functions of the discrete independent variable $i$, and analyticity and asymptoticity properties of the Jost solutions as functions of the spectral parameter $\zeta$.

\begin{definition}
The functions $\lam,\rlam$ and $\up,\rup$ are defined by
\begin{subequations}
\begin{align}
&\vp(i;\zeta)=\lam(i;\zeta)\prod_{r=0}^{i-1}\left(\frac{\pz(r)+\zeta}{\pz(r)-\zeta}\right)^{\frac12}, \;\; \rvp(i;\zeta)=\rlam(i;\zeta)\prod_{r=0}^{i-1}\left(\frac{\pz(r)-\zeta}{\pz(r)+\zeta}\right)^{\frac12} \\
&\psi(i;\zeta)=\up(i;\zeta)\prod_{r=0}^{i-1}\left(\frac{\pz(r)-\zeta}{\pz(r)+\zeta}\right)^{\frac12}, \;\; \rpsi(i;\zeta)=\rup(i;\zeta)\prod_{r=0}^{i-1}\left(\frac{\pz(r)+\zeta}{\pz(r)-\zeta}\right)^{\frac12}.
\end{align}
\end{subequations}
\end{definition}

\begin{proposition}\label{prop:jostsum}
For $\zeta\neq0$ the functions $\lam$ and $\up$ satisfy the following summation equations:
\begin{align}\label{eq:lamsum}
&\lam(i;\zeta)=1+\frac1{2\zeta}\sum_{l=-\infty}^{i-1}\left[1-\prod_{r=l}^{i-1}\left(\frac{\pz(r)-\zeta}{\pz(r)+\zeta}\right)\right]\pot(l)\lam(l;\zeta) \\
\label{eq:upsum}
&\up(i;\zeta)=1+\frac1{2\zeta}\sum_{l=i+1}^{+\infty}\left[1-\prod_{r=i}^{l-1}\left(\frac{\pz(r)-\zeta}{\pz(r)+\zeta}\right)\right]\pot(l)\up(l;\zeta).
\end{align}
\begin{proof}
Equation \eqref{eq:ds11} for $\lam$ becomes
\[
(\po+\zeta)\lamoo-(\pz+\po)\lamo+(\pz-\zeta)\lam=\poto\lamo
\]
which in terms of $i$ may be written as
\begin{align*}
&\pz(i+1)\Bigl[\lam(i+2;\zeta)-\lam(i+1;\zeta)\Bigr]-\pz(i)\Bigl[\lam(i+1;\zeta)-\lam(i;\zeta)\Bigr]+\zeta\Bigl[\lam(i+2;\zeta)-\lam(i;\zeta)\Bigr]=\pot(i+1)\lam(i+1;\zeta),
\end{align*}
and after summing from $l=-\infty$ to $l=i-1$, and using $\lam\to1$ as $i\to-\infty$, we have
\[
\lam(i+1;\zeta)\Bigl[\pz(i)+\zeta\Bigr]-\lam(i;\zeta)\Bigl[\pz(i)-\zeta\Bigr]=2\zeta+\sum_{l=-\infty}^{i}\pot(l)\lam(l;\zeta).
\]
We now multiply this equation by the summing factor $s(i):=\prod_{r=0}^{i-1}\left(\frac{\pz(r)+\zeta}{\pz(r)-\zeta}\right)$, which gives
\begin{align*}
&\lam(i+1;\zeta)s(i+1)-\lam(i;\zeta)s(i)=\Bigl[s(i+1)-s(i)\Bigl]+\frac1{2\zeta}\Bigl[s(i+1)-s(i)\Bigl]\sum_{l=-\infty}^{i}\pot(l)\lam(l;\zeta)
\end{align*}
and then we sum from $j=i_o\leq i-1$ to $j=i-1$, obtaining
\begin{align*}
&\lam(i;\zeta)s(i)-\lam(i_o;\zeta)s(i_o)=\Bigl[s(i)-s(i_o)\Bigl]+\frac1{2\zeta}\sum_{j=i_o}^{i-1}\Bigl[s(j+1)-s(j)\Bigl]\sum_{l=-\infty}^{j}\pot(l)\lam(l;\zeta).
\end{align*}
We now let $i_0\to-\infty$ and assume that $s(i)\to0$ as $i\to-\infty$. By changing the order of summation the double sum can be rewritten as
\begin{align*}
\sum_{j=-\infty}^{i-1}\sum_{l=-\infty}^{j}\Bigl[s(j+1)-s(j)\Bigl]\pot(l)\lam(l;\zeta)&=\sum_{l=-\infty}^{i-1}\pot(l)\lam(l;\zeta)\sum_{j=l}^{i-1}\Bigl[s(j+1)-s(j)\Bigl] \\
&=\sum_{l=-\infty}^{i-1}\Bigl[s(i)-s(l)\Bigl]\pot(l)\lam(l;\zeta)
\end{align*}
and so the summation equation becomes
\begin{align*}
\lam(i;\zeta)s(i)&=s(i)+\frac1{2\zeta}\sum_{l=-\infty}^{i-1}\Bigl[s(i)-s(l)\Bigl]\pot(l)\lam(l;\zeta) \\
\Rightarrow \;\;\lam(i;\zeta)&=1+\frac1{2\zeta}\sum_{l=-\infty}^{i-1}\left[1-\prod_{r=l}^{i-1}\left(\frac{\pz(r)-\zeta}{\pz(r)+\zeta}\right)\right]\pot(l)\lam(l;\zeta),
\end{align*}
which is equation \eqref{eq:lamsum}. Equation \eqref{eq:upsum} follows in a similar manner by summing to $i=+\infty$ and using $\up\to1$ as $i\to+\infty$. 
\end{proof}
\end{proposition}

\begin{proposition}
At $\zeta=0$ the functions $\lam$ and $\up$ satisfy the following summation equations:
\begin{align}\label{eq:lamsum0}
&\lam(i;0)=1+\sum_{l=-\infty}^{i-1}\left[\sum_{j=l}^{i-1}\frac1{\pz(j)}\right]\pot(l)\lam(l;0) \\
\label{eq:upsum0}
&\up(i;0)=1+\sum_{l=i+1}^{+\infty}\left[\sum_{j=i}^{l-1}\frac1{\pz(j)}\right]\pot(l)\up(l;0).
\end{align}
\begin{proof}
At $\zeta=0$, equation \eqref{eq:ds11} for $\lam$ becomes
\begin{align*}
&\pz(i+1)\Bigl[\lam(i+2;0)-\lam(i+1;0)\Bigr]-\pz(i)\Bigl[\lam(i+1;0)-\lam(i;0)\Bigr]=\pot(i+1)\lam(i+1;0),
\end{align*}
which after summing from $l=-\infty$ to $l=i-1$ gives
\[
\pz(i)\Bigl[\lam(i+1;0)-\lam(i;0)\Bigr]=\sum_{l=-\infty}^i\pot(l)\lam(l;0).
\]
After dividing through by $\pz(i)$ and summing again from $j=-\infty$ to $j=i-1$ we obtain
\[
\lam(i;0)=1+\sum_{j=-\infty}^{i-1}\sum_{l=-\infty}^{j}\left[\frac1{\pz(j)}\right]\pot(l)\lam(l;0),
\]
and by changing the order of summation we have
\[
\lam(i;0)=1+\sum_{l=-\infty}^{i-1}\left[\sum_{j=l}^{i-1}\frac1{\pz(j)}\right]\pot(l)\lam(l;0)
\]
which is equation \eqref{eq:lamsum}. The result \eqref{eq:upsum} follows in a similar manner.
\end{proof}
\end{proposition}

\begin{proposition}
For $\zeta\neq0$ the summation equations \eqref{eq:lamsum} and \eqref{eq:upsum} have the Neumann series solutions
\begin{equation}\label{eq:lamupseries}
\lam(i;\zeta)=\sum_{k=0}^{+\infty}\frac{H_k(i;\zeta)}{\zeta^k}, \;\;\;\;\; \up(i;\zeta)=\sum_{k=0}^{+\infty}\frac{J_k(i;\zeta)}{\zeta^k}
\end{equation}
where
\begin{align}\label{eq:Hkrr}
&H_0=1, \;\;\;\;\; H_{k+1}(i;\zeta)=\frac12\sum_{l=-\infty}^{i-1}\left[1-\prod_{r=l}^{i-1}\left(\frac{\pz(r)-\zeta}{\pz(r)+\zeta}\right)\right]\pot(l)H_k(l;\zeta), \\
\label{eq:Jkrr}
&J_0=1, \;\;\;\;\; J_{k+1}(i;\zeta)=\frac12\sum_{l=i+1}^{+\infty}\left[1-\prod_{r=i}^{l-1}\left(\frac{\pz(r)-\zeta}{\pz(r)+\zeta}\right)\right]\pot(l)J_k(l;\zeta).
\end{align}
\begin{proof}
Inserting this series expression for $\lam$ into the summation equation \eqref{eq:lamsum} gives
\begin{align*}
\lam(i;\zeta)&=1+\frac1{2\zeta}\sum_{l=-\infty}^{i-1}\left[1-\prod_{r=l}^{i-1}\left(\frac{\pz(r)-\zeta}{\pz(r)+\zeta}\right)\right]\pot(l)\left(\sum_{k=0}^{+\infty}\frac{H_k(l;\zeta)}{\zeta^k}\right) \\
&=1+\sum_{k=0}^{+\infty}\frac1{\zeta^{k+1}}\left(\frac12\sum_{l=-\infty}^{i-1}\left[1-\prod_{r=l}^{i-1}\left(\frac{\pz(r)-\zeta}{\pz(r)+\zeta}\right)\right]\pot(l)H_k(l;\zeta)\right) \\
&=1+\sum_{k=0}^{+\infty}\frac{H_{k+1}(i;\zeta)}{\zeta^{k+1}} \\
&=\sum_{k=0}^{+\infty}\frac{H_k(i;\zeta)}{\zeta^k}
\end{align*}
as required. The proof for $\up$ follows in a similar fashion.
\end{proof}
\end{proposition}

\begin{proposition}
At $\zeta=0$ the summation equations \eqref{eq:lamsum0} and \eqref{eq:upsum0} have the Neumann series solutions
\begin{equation}\label{eq:lamupseries0}
\lam(i;0)=\sum_{k=0}^{+\infty}H_k^o(i), \;\;\;\;\; \up(i;0)=\sum_{k=0}^{+\infty}J_k^o(i)
\end{equation}
where
\begin{align}\label{eq:Hkorr}
&H_0^o=1, \;\;\;\;\; H_{k+1}^o(i)=\sum_{l=-\infty}^{i-1}\left[\sum_{j=l}^{i-1}\frac1{\pz(j)}\right]\pot(l)H_k^o(l), \\
\label{eq:Jkorr}
&J_0^o=1, \;\;\;\;\; J_{k+1}^o(i)=\sum_{l=i+1}^{+\infty}\left[\sum_{j=i}^{l-1}\frac1{\pz(j)}\right]\pot(l)J_k^o(l).
\end{align}
\begin{proof}
Inserting this series expression for $\lam$ into the summation equation \eqref{eq:lamsum0} gives
\begin{align*}
\lam(i;0)&=1+\sum_{l=-\infty}^{i-1}\left[\sum_{j=l}^{i-1}\frac1{\pz(j)}\right]\pot(l)\left(\sum_{k=0}^{+\infty}H_k^o(l)\right) \\
&=1+\sum_{k=0}^{+\infty}\left(\sum_{l=-\infty}^{i-1}\left[\sum_{j=l}^{i-1}\frac1{\pz(j)}\right]\pot(l)H_k^o(l)\right) \\
&=1+\sum_{k=0}^{+\infty}H_{k+1}^o(i) \\
&=\sum_{k=0}^{+\infty}H_k^o(i)
\end{align*}
as required. The proof for $\up$ is similar. 
\end{proof}
\end{proposition}

\begin{theorem}\label{th:jostanal1}
Assume that 
\begin{equation}\label{eq:potsum}
\sum_{i=-\infty}^{+\infty}|\pot(i)|(1+|i|)<\infty,
\end{equation}
and that $\pz(r)>0$ for all $r\in\cI$. Let $\cR^+$ denote the half-plane 
\begin{equation}\label{eq:R+}
\cR^+:=\bigl\{\;\zeta\;:\;{\rm Re}(\zeta)\geq0\bigr\}.
\end{equation}
Then for $\zeta\in\cR^+$,
\begin{subequations}
\begin{align}\label{eq:lamest1}
&|\lam(i;\zeta)-1|\leq C_1 \;\;\; {\rm for} \;\; \zeta\neq0 \\
\label{eq:lamest10}
&|\lam(i;\zeta)-1|\leq C_2(1+\max\{0,i\}) \\
\label{eq:upest1}
&|\up(i;\zeta)-1|\leq C_3 \;\;\; {\rm for} \;\; \zeta\neq0 \\
\label{eq:upest10}
&|\up(i;\zeta)-1|\leq C_4(1+\max\{0,-i\}) 
\end{align}
\end{subequations}
where $C_1\to C_4$ are constants. For all $\zeta\in\cR^+$ the series solutions for $\lam$ and $\up$ converge absolutely in $i$, and uniformly if $\zeta\neq0$. For each $i$, $\lam$ and $\up$ are continuous functions of $\zeta$ in $\cR^+$, and analytic functions of $\zeta$ in the interior of this half-plane. 
\begin{proof}
The proof of this Theorem is obtained by showing absolute and uniform convergence of the Neumann series representation of the Jost solutions. The complete details, reminiscent of the analysis given in \cite{dt:79} for the continuous case, are given in the Appendix. The estimates obtained agree with those obtained in \cite{bpps:01}, \cite{s:02} and \cite{bj:10} for the a similar spectral problem.
\end{proof}
\end{theorem}

\begin{corollary}\label{cor:jostasymp}
For $\zeta\in\cR^+$ the functions $\lam$ and $\up$ have the following asymptotic behaviour:
\begin{subequations}\label{eq:jostasymp}
\begin{align}
&\lam(i;\zeta)=1+\cO\left(\frac1{\zeta}\right) \;\;\;\;\; {\rm as} \;\; |\zeta|\to\infty \\
&\up(i;\zeta)=1+\cO\left(\frac1{\zeta}\right) \;\;\;\;\; {\rm as} \;\; |\zeta|\to\infty.
\end{align}
\end{subequations}
\begin{proof}
From the series solution of $\lam$ we have
\[
\lam(i;\zeta)=1+\sum_{k=1}^{+\infty}\frac{H_k(i;\zeta)}{\zeta^k}.
\]
For $\zeta\in\cR^+$, $\zeta\neq0$ however, from the Appendix we have that $|H_k|\leq K$ for some constant $K$, and thus $H_k=\cO(1)$ as $|\zeta|\to\infty$, for all $k\geq1$. This proves the result, and a similar argument works for the series solution for $\up$.
\end{proof}\end{corollary}

\begin{corollary}
Theorem \ref{th:jostanal1} and Corollary \ref{cor:jostasymp} hold for the functions $\rlam$ and $\rup$ for $\zeta$ in the half-plane
\[
\cR^-:=\bigl\{\zeta:{\rm Re}(\zeta)\leq0\bigr\}.
\]
\begin{proof}
This follows from the fact that $\rlam(i;\zeta)=\lam(i;-\zeta)$ and $\rup(i;\zeta)=\up(i;-\zeta)$.
\end{proof}
\end{corollary}

\subsection{Analyticity and Asymptoticity Properties of $\az$ and $\bz$}

We now look at analyticity and asymptoticity properties of $\az=\az(\zeta)$ and $\bz=\bz(\zeta)$, which are defined by equation \eqref{eq:ab}. These are related to the reflection coefficient $R$ and transmission coefficient $T$ by
\[
R=\frac{\bz}{\az}, \;\;\;\;\; T=\frac1{\az}.
\]

\begin{proposition}\label{prop:abanal}
$\az$ and $\bz$ have the following properties:
\begin{enumerate}
\item[-] $\az$ is analytic in the interior of $\cR^+$ and continuous in $\cR^+$, except possibly at $\zeta=0$
\item[-] $\bz$ is continuous on the imaginary $\zeta$-axis, except possibly at $\zeta=0$.
\end{enumerate}
\begin{proof}
Taking the Wronskian of $\psi=\az\rvp+\bz\vp$ we have
\begin{align*}
\az(\zeta)&=\frac1{2\zeta}W(\psi,\vp)=\frac1{2\zeta}\Bigl((\pz(i)+\zeta)\lam(i+i;\zeta)\up(i;\zeta)-(\pz(i)-\zeta)\lam(i;\zeta)\up(i+1;\zeta)\Bigr) \\
\bz(\zeta)&=\frac1{2\zeta}W(\rvp,\vp)=\left(\frac{\pz(i)-\zeta}{2\zeta}\right)\prod_{r=0}^{i-1}\left(\frac{\pz(r)-\zeta}{\pz(r)+\zeta}\right)\Bigl(\rlam(i;\zeta)\up(i+1;\zeta)-\rlam(i+1;\zeta)\up(i;\zeta)\Bigr)
\end{align*}
Since $\lam$ and $\up$ are continuous in $\cR^+$ and analytic in the interior of this region, $\az$ also has this property, except possibly at $\zeta=0$. The expression for $\bz$ however is only valid on the intersection of $\cR^+$ and $\cR^-$, i.e. the imaginary $\zeta$-axis. Since $\rlam$ and $\up$ are continuous here, $\bz$ also has this property, except possibly at $\zeta=0$. 
\end{proof}
\end{proposition}

\begin{proposition}\label{prop:absum}
For $\zeta\neq0$ the functions $\az$ and $\bz$ can be expressed as
\begin{align}\label{eq:asum}
\az(\zeta)&=1+\frac1{2\zeta}\sum_{l=-\infty}^{+\infty}\pot(l)\xx\up(l;\zeta) \\
\label{eq:bsum}
\bz(\zeta)&=\frac1{2\zeta}\sum_{l=-\infty}^{+\infty}\left[\prod_{r=0}^{l-1}\left(\frac{\pz(r)-\zeta}{\pz(r)+\zeta}\right)\right]\pot(l)\xx\up(l;\zeta).
\end{align}
\begin{proof}
The summation equation \eqref{eq:upsum} for $\up$ may be written as
\begin{align}\label{eq:upsum2}
\up(i;\zeta)&=\left(\xx1+\frac1{2\zeta}\sum_{l=i+1}^{+\infty}\pot(l)\up(l;\zeta)\xx\right)+\prod_{r=0}^{i-1}\left(\frac{\pz(r)-\zeta}{\pz(r)+\zeta}\right)\left(\xx\frac1{2\zeta}\sum_{l=i+1}^{+\infty}\left[\prod_{r=0}^{l-1}\left(\frac{\pz(r)-\zeta}{\pz(r)+\zeta}\right)\right]\pot(l)\xx\up(l;\zeta)\xx\right).
\end{align}
Taking the limit $i\to-\infty$ and comparing this with
\[
\up(i;\zeta)\sim\az(\zeta)+\bz(\zeta)\prod_{r=0}^{i-1}\left(\frac{\pz(r)-\zeta}{\pz(r)+\zeta}\right) \;\; {\rm as} \;\;\; i\to-\infty
\]
gives the desired result.
\end{proof}
\end{proposition}

\begin{proposition}\label{prop:abasymp}
For $\zeta\in\cR^+$ we have
\begin{equation}\label{eq:aasymp}
\az(\zeta)=1+\cO\left(\frac1{\zeta}\right) \;\; {\rm as} \;\;\; |\zeta|\to\infty
\end{equation}
and for $\zeta$ on the imaginary axis we have
\begin{equation}\label{eq:basymp}
\bz(\zeta)=\cO\left(\frac1{\zeta}\right) \;\; {\rm as} \;\;\; |\zeta|\to\infty.
\end{equation}
\begin{proof}
Inserting the asymptotic behaviour \eqref{eq:jostasymp} of $\up$ into the expression \eqref{eq:asum} for $\az$ gives
\[
\az(\zeta)=1+\frac1{2\zeta}\sum_{l=-\infty}^{+\infty}\pot(l)\xx\left[1+\cO\left(\frac1{\zeta}\right)\right]
\]
and since
\[
\left|\sum_{l=-\infty}^{+\infty}\pot(l)\right|\leq\sum_{l=-\infty}^{+\infty}|\pot(l)|<\infty
\]
this proves \eqref{eq:aasymp}. Performing the same task for the expression \eqref{eq:bsum} for $\bz$ gives
\[
\bz(\zeta)=\frac1{2\zeta}\sum_{l=-\infty}^{+\infty}\left[\prod_{r=0}^{l-1}\left(\frac{\pz(r)-\zeta}{\pz(r)+\zeta}\right)\right]\pot(l)\xx\left[1+\cO\left(\frac1\zeta\right)\right],
\]
and since for purely imaginary $\zeta$ we have
\[
\left|\sum_{l=-\infty}^{+\infty}\left[\prod_{r=0}^{l-1}\left(\frac{\pz(r)-\zeta}{\pz(r)+\zeta}\right)\right]\pot(l)\right|\leq\sum_{l=-\infty}^{+\infty}|\pot(l)|<\infty,
\]
this proves \eqref{eq:basymp}.
\end{proof}
\end{proposition}

\begin{theorem}\label{thm:a1}
The function $\az$ has a finite number of bounded isolated zeroes $\bigl\{\zeta_k,\xx k=1,...,M\bigr\}$ in the interior of $\cR^+$, and moreover every $\zeta_k$ is purely real and satisfies $\zeta_k\leq p_r$ for all parameters $p_r$ existing along $\Gamma$. At each zero of $\az$ we have $\psi(i;\zeta_k)=\bz(\zeta_k)\vp(i;\zeta_k)$, and
\begin{equation}\label{eq:normconst}
\sum_{i=-\infty}^{+\infty}\left(\frac{\vp(i;\zeta_k)\vp(i+1;\zeta_k)}{(\pz(i)^2-\zeta_k^2)^{\frac12}}\right)=\frac{\az'(\zeta_k)}{\bz(\zeta_k)},
\end{equation}
where $\az'(\zeta)$ denotes the derivative of $\az$ with respect to $\zeta$.
\begin{proof}
Since $\az\sim1$ as $|\zeta|\to\infty$ it follows that there exists some constant $C_o$ such that $|\zeta_k|<C_o$ for every $k$. Since $2\zeta\az(\zeta)=W(\psi,\vp)$ it follows that for every $k$, $\vp(i;\zeta_k)$ and $\psi(i;\zeta_k)$ are linearly dependent, so we may write $\psi(i;\zeta_k)=b_k\vp(i;\zeta_k)$ for some constant $b_k$. This implies that
\[
\up(i;\zeta_k)\sim b_k\prod_{r=0}^{i-1}\left(\frac{\pz(r)-\zeta_k}{\pz(r)+\zeta_k}\right) \;\; {\rm as} \;\;\; i\to-\infty,
\]
and so by equation \eqref{eq:upsum2} we have
\begin{align*}
&\zeta_k=-\frac12\sum_{l=-\infty}^{+\infty}\pot(l)\xx\up(l;\zeta_k) \\
&b_k=\frac1{2\zeta_k}\sum_{l=-\infty}^{+\infty}\left[\prod_{r=0}^{l-1}\left(\frac{\pz(r)-\zeta_k}{\pz(r)+\zeta_k}\right)\right]\pot(l)\xx\up(l;\zeta_k).
\end{align*}
Thus $b_k=\bz(\zeta_k)$ for every $k$. Now consider the scattering problem \eqref{eq:ds11} for $\vp$ at $\zeta=\zeta_k$:
\[
(\po^2-\zeta_k^2)^{\frac12}\xx\vp(i+2;\zeta_k)-\bigl(\xx\pz+\po+\poto\xx\bigr)\xx\vp(i+1;\zeta_k)+(\pz^2-\zeta_k^2)^{\frac12}\xx\vp(i;\zeta_k)=0.
\]
For every $\zeta_k$ in the interior of $\cR^+$ we have
\begin{align*}
\vp(i;\zeta_k)&\sim\prod_{r=0}^{i-1}\left(\frac{\pz(r)+\zeta_k}{\pz(r)-\zeta_k}\right)^{\frac12}\to0 \;\; {\rm as} \;\; i\to-\infty, \\
\vp(i;\zeta_k)&\sim\bz(\zeta_k)\prod_{r=0}^{i-1}\left(\frac{\pz(r)-\zeta_k}{\pz(r)+\zeta_k}\right)^{\frac12}\to0 \;\; {\rm as} \;\; i\to+\infty,
\end{align*}
and thus $\vp(i;\zeta_k)$ is summable over all $i$. If we multiply the scattering problem for $\vp(i;\zeta_k)$ by $\vp^*(i+1;\zeta_k^*)$, sum over all $i$ and define
\[
s(i;\zeta_k):=\vp(i+1;\zeta_k)\vp^*(i;\zeta_k^*)+\vp^*(i+1;\zeta_k^*)\vp(i;\zeta_k)\in\mathbb{R}
\]
then we have
\begin{align*}
\sum_{i=-\infty}^{+\infty}(\pz(i)^2-\zeta_k^2)^{\frac12}s(i;\zeta_k)=\sum_{i=-\infty}^{+\infty}\Bigl[\pz(i)+\pz(i+1)+\pot(i+1)\Bigr]|\vp(i+1;\zeta_k)|^2,
\end{align*}
which implies that $\zeta_k$ must be real and that $0<\zeta_k\leq \pz(i)$ for every $i$. Thus $\zeta_k$ must be less than every parameter $p_r$ through which $\pz(i)$ cycles, which proves the given statement. 
\par
Now $\az$ has isolated zeroes along the positive real $\zeta$ axis which are all bounded. The only way that there could be an infinite number of these zeroes is if they formed a limiting sequence which accumulated at $\zeta=0$. We will show that this is not possible. Suppose that such a sequence $\{\zeta_k\}$ of zeroes exists: $\lim_{k\to\infty}\zeta_k=0$. Then at each $\zeta_k$ we have
\[
\bz(\zeta_k)=\frac{\psi(i;\zeta_k)}{\vp(i;\zeta_k)},
\]
and so
\[
\lim_{k\to\infty}|\bz(\zeta_k)-\bz(0)|=\lim_{k\to\infty}\left|\frac{\psi(i;\zeta_k)}{\vp(i;\zeta_k)}-\frac{\psi(i;0)}{\vp(i;0)}\right|=0
\]
since the Jost solutions are continuous at $\zeta=0$. At $\zeta=0$ however we have $\vp(i;0)=\rvp(i;0)$ and so
\[
\az(0)+\bz(0)=\left(\frac{\psi(i;0)}{\vp(i;0)}\right)=\lim_{k\to\infty}\left(\frac{\psi(i;\zeta_k)}{\vp(i;\zeta_k)}\right)=\lim_{k\to\infty}\bz(\zeta_k)=\bz(0)
\]
which implies that $\az(0)=0$, which in turn contradicts Proposition \ref{prop:aneq0}. Thus $\az$ has only a finite number of zeroes in $\cR^+$. 
\par
Finally to prove \eqref{eq:normconst} we define the following two useful functions:
\[
W_\vp(i;\zeta):=W(\vp,\vp'), \;\;\;\;\; W_\psi(i;\zeta):=W(\psi,\psi').
\]
We differentiate $2\zeta\az=W(\psi,\vp)$ to obtain
\begin{align}\label{eq:a'}
2\zeta_k\az'(\zeta_k)&=W(\psi(i;\zeta_k),\vp'(i;\zeta_k))+W(\psi(i;\zeta_k),\vp'(i;\zeta_k))=\bz(\zeta_k)W_\vp(i;\zeta_k)-\frac1{\bz(\zeta_k)}W_\psi(i;\zeta_k).
\end{align}
Now consider the difference of two equations: firstly the derivative of the scattering problem \eqref{eq:ds11} for $\vp$ multiplied by $\vp(i+1;\zeta)$, and secondly the (un-differentiated) scattering problem for $\vp$ multiplied by $\vp'(i+1;\zeta)$. This gives
\[
W_\vp(i+1;\zeta)-W_\vp(i;\zeta)=\zeta\left[\frac{\vp(i;\zeta)\vp(i+1;\zeta)}{(\pz(i)^2-\zeta^2)^{\frac12}}+\frac{\vp(i+1;\zeta)\vp(i+2;\zeta)}{(\pz(i+1)^2-\zeta^2)^{\frac12}}\right],
\]
which may be summed to give
\[
W_\vp(i;\zeta)=\zeta\sum_{l=-\infty}^{i-1}\left[\frac{\vp(l;\zeta)\vp(l+1;\zeta)}{(\pz(l)^2-\zeta^2)^{\frac12}}+\frac{\vp(l+1;\zeta)\vp(l+2;\zeta)}{(\pz(l+1)-\zeta^2)^{\frac12}}\right].
\]
One can then perform the same task with $\psi$, only instead this time summing from $i$ to $+\infty$, to obtain
\[
W_\psi(i;\zeta)=-\zeta\sum_{l=i}^{+\infty}\left[\frac{\psi(l;\zeta)\psi(l+1;\zeta)}{(\pz(l)^2-\zeta^2)^{\frac12}}+\frac{\psi(l+1;\zeta)\psi(l+2;\zeta)}{(\pz(l+1)-\zeta^2)^{\frac12}}\right].
\]
Now set $\zeta=\zeta_k$ and rewrite $\psi(i;\zeta_k)=\bz(\zeta_k)\vp(i;\zeta_k)$. The expression \eqref{eq:a'} then becomes
\begin{align*}
2\zeta_k\az'(\zeta_k)&=\zeta_k\bz(\zeta_k)\sum_{l=-\infty}^{+\infty}\left[\frac{\vp(l;\zeta_k)\vp(l+1;\zeta_k)}{(\pz(l)^2-\zeta_k^2)^{\frac12}}+\frac{\vp(l+1;\zeta_k)\vp(l+2;\zeta_k)}{(\pz(l+1)-\zeta_k^2)^{\frac12}}\right] \\
&=2\zeta_k\bz(\zeta_k)\sum_{l=-\infty}^{+\infty}\left[\frac{\vp(l;\zeta_k)\vp(l+1;\zeta_k)}{(\pz(l)^2-\zeta_k^2)^{\frac12}}\right],
\end{align*}
which gives \eqref{eq:normconst}.
\end{proof}
\end{theorem}

The sum in equation \eqref{eq:normconst} is of fundamental importance to the scattering problem, and as such we make the following definition.

\begin{definition}\label{def:normconst}
The {\bf square eigenfunction} $\Phi$ is defined to be
\begin{equation}\label{eq:sqe}
\Phi(i;\zeta):=\frac{\vp(i;\zeta)\vp(i+1;\zeta)}{(\pz(i)^2-\zeta^2)^{\frac12}},
\end{equation}
and we define the {\bf normalisation constants} $\{\cz_k,\xx k=1,...,M\}$ to be
\begin{equation}\label{eq:nc} 
\cz_k:=\frac{\bz(\zeta_k)}{\az'(\zeta_k)}=\left[\sum_{i=-\infty}^{+\infty}\Phi(i;\zeta_k)\right]^{-1}.
\end{equation}
\end{definition}

From this definition it is clear that we have the following result

\begin{theorem}\label{th:a2}
If the normalisation constants $\cz_k$ are all finite, then every zero $\az$ in $\cR^+$ is simple.
\end{theorem}

One possible way of ensuring that all normalisation constants are finite is by using the result
\begin{align*}
\frac1{\cz_k}&=\sum_{i=-\infty}^{+\infty}\left(\frac{\vp(i;\zeta_k)\vp(i+1;\zeta_k)}{(\pz(i)^2-\zeta_k^2)^{\frac12}}\right)=\sum_{i=-\infty}^{+\infty}\bigl[\pz(i)+\pz(i+1)+\pot(i+1)\bigr]|\vp(i;\zeta_k)|^2.
\end{align*}
Thus if we impose, for example, that $\bigl[\pz(i)+\pz(i+1)+\pot(i+1)\bigr]>0$ for all $i$, then every $\cz_k$ will be finite. We do not dwell on this further, but assume henceforth that all normalisation constants are finite. 
\par
This concludes the forward scattering of $\phi_1$. The main results of this section are the construction of the Jost solutions $\lam(i;\zeta), \up(i;\zeta)$, the spectral functions $\az(\zeta), \bz(\zeta)$, and the knowledge of their analyticity and asymptoticity properties.

\section{``Time" Evolution of the Scattering Data}\label{sec:timeevo}

We now consider how the spectral functions $\az$, $\bz$ and the normalisation constants $\cz_k$ depend on the N lattice variables. As an analogy to the continuous theory, this is the calculation of the ``time" dependence of these functions, with respect to the arbitrary number of discrete ``time" variables. 
\par
From the Lax pair the equation governing the evolution of the Jost solutions (obtained by eliminating the second component) in any one particular lattice direction with variable $n_k$ and parameter $p_k$ is given by
\begin{equation}\label{eq:nkevolution}
(p_k^2-\zeta^2)^{\frac12}\asf{\asf\vp}-\left(\frac{P_k(\asf{\asf u}-u)}{\asf{\cU}}\right)\asf\vp+(p_k^2-\zeta^2)^{\frac12}\vp=0,
\end{equation}
where $\asf{u}$ denotes an iteration of $u$ in the $n_k$-direction. From the boundary conditions \eqref{eq:uBC} for $u$, for all $k=1,...,\rN$ we have
\begin{equation}\label{eq:nkboundary}
\frac{P_k(\asf{\asf u}-u)}{\asf{\cU}}\sim2p_k \;\;\; {\rm as} \;\; i\to\pm\infty.
\end{equation}
In other words at both ends of the staircase the dependence of the Jost solutions on the lattice variable $n_k$ is governed by
\begin{equation}\label{eq:nkevolution2}
(p_k^2-\zeta^2)^{\frac12}\asf{\asf\vp}-2p_k\asf\vp+(p_k^2-\zeta^2)^{\frac12}\vp=0.
\end{equation}

\subsection{``Staircase Directions" ($\in\cI$)} $\\$ 

Recall that $\cI$ denotes the set of lattice directions in which the staircase of initial values $\Gamma$ iterates. Due to the way it was defined in Section \ref{sec:motivation}, the set of initial values along $\Gamma$ defines a well-posed initial value problem \cite{av:04} for all points in each of the ``staircase directions" indexed by $\cI$. For example given the staircase of initial values in Figure \ref{fig:staircasetimeevo1}, all points in this plane are uniquely defined by \eqref{eq:Q3}. Likewise in Figure \ref{fig:staircasetimeevo3}, given this staircase of initial values, all points in this three-dimensional space are uniquely defined by \eqref{eq:Q3}.
\begin{figure}[h!]
\begin{minipage}{0.25\textwidth}
\begin{tikzpicture}[scale=0.7]
\draw  (-0.3,0) -- (4.3,0);
\draw  (-0.3,4) -- (4.3,4);
\draw  (0,-0.3) -- (0,4.3);
\draw  (4,-0.3) -- (4,4.3);
\draw  (-0.3,1) -- (4.3,1);
\draw  (-0.3,3) -- (4.3,3);
\draw  (1,-0.3) -- (1,4.3);
\draw  (3,-0.3) -- (3,4.3);
\draw  (2,-0.3) -- (2,4.3);
\draw  (-0.3,2) -- (4.3,2);
\draw [ultra thick, <->] (0,-0.3) -- (0,0) -- (1,0) -- (1,1) -- (2,1) -- (2,2) -- (3,2) -- (3,3) -- (4,3) -- (4,4) -- (4.3,4);
\node [below right] at (2,2) {$\Gamma$};
\end{tikzpicture}
\subcaption{All points are uniquely determined by $\Gamma$}
\label{fig:staircasetimeevo1}
\end{minipage}
\hspace{0.5cm}
\begin{minipage}{0.25\textwidth}
\begin{tikzpicture}[scale=0.7]
\draw  (0,0) -- (0.7,0.7) -- (0.7,1.7) -- (0,2.4) -- (-0.7,1.7) -- (-0.7,0.7) -- (0,0) -- (0,1) -- (0.7,1.7) -- (0,1) -- (-0.7,1.7);
\draw  (1.4,0) -- (2.1,0.7) -- (2.1,1.7) -- (1.4,2.4) -- (0.7,1.7) -- (0.7,0.7) -- (1.4,0) -- (1.4,1) -- (2.1,1.7) -- (1.4,1) -- (0.7,1.7);
\draw  (2.8,0) -- (3.5,0.7) -- (3.5,1.7) -- (2.8,2.4) -- (2.1,1.7) -- (2.1,0.7) -- (2.8,0) -- (2.8,1) -- (3.5,1.7) -- (2.8,1) -- (2.1,1.7);
\draw  (-0.7,-1.7) -- (0,-1) -- (0,0) -- (-0.7,0.7) -- (-1.4,0) -- (-1.4,-1) -- (-0.7,-1.7) -- (-0.7,-0.7) -- (0,0) -- (-0.7,-0.7) -- (-1.4,0);
\draw  (0.7,-1.7) -- (1.4,-1) -- (1.4,0) -- (0.7,0.7) -- (0,0) -- (0,-1) -- (0.7,-1.7) -- (0.7,-0.7) -- (1.4,0) -- (0.7,-0.7) -- (0,0);
\draw  (2.1,-1.7) -- (2.8,-1) -- (2.8,0) -- (2.1,0.7) -- (1.4,0) -- (1.4,-1) -- (2.1,-1.7) -- (2.1,-0.7) -- (2.8,0) -- (2.1,-0.7) -- (1.4,0);
\draw  (3.5,-1.7) -- (4.2,-1) -- (4.2,0) -- (3.5,0.7) -- (2.8,0) -- (2.8,-1) -- (3.5,-1.7) -- (3.5,-0.7) -- (4.2,0) -- (3.5,-0.7) -- (2.8,0);
\draw [ultra thick, <->] (-1.6,-0.8) -- (-1.4,-1) -- (-1.4,0) -- (-0.7,0.7) -- (0,0) -- (0.7,0.7) -- (1.4,0) -- (1.4,1) -- (2.1,1.7) -- (2.8,1) -- (3.5,1.7) -- (4.2,1) -- (4.2,1.4);
\node [above left] at (-0.9,0.4) {\small $\Gamma$};
\end{tikzpicture}
\subcaption{All points are uniquely determined by $\Gamma$}
\label{fig:staircasetimeevo3}
\end{minipage}
\hspace{0.5cm}
\begin{minipage}{0.25\textwidth}
\begin{tikzpicture}[scale=0.7]
\draw  (-0.3,-0.3) -- (2.3,2.3);
\draw  (0.7,-0.3) -- (3.3,2.3);
\draw  (1.7,-0.3) -- (4.3,2.3);
\draw  (2.7,-0.3) -- (5.3,2.3);
\draw (3.7,-0.3) -- (6.3,2.3);
\draw (-0.3,0) -- (4.3,0);
\draw (0.2,0.5) -- (4.8,0.5);
\draw (0.7,1) -- (5.3,1);
\draw (1.3,1.5) -- (5.8,1.5);
\draw (1.8,2) -- (6.3,2);
\draw [->] (0,-0.3) -- (0,3.3);
\draw [->] (2,1.7) -- (2,5.3);
\draw [ultra thick, <->] (-0.3,-0.3) -- (0,0) -- (1,0) -- (1.5,0.5) -- (2.5,0.5) -- (3,1) -- (4,1) -- (4.5,1.5) -- (5.5,1.5) -- (6,2) -- (6.3,2);
\node [below right] at (2.85,1.05) {{\small $\Gamma$}};
\node [left] at (0,2) {$\notin\cI$};
\node [right] at (5.3,1) {$\in\cI$};
\end{tikzpicture}
\subcaption{Points in the plane are uniquely determined but those in the vertical direction are not}
\label{fig:staircasetimeevo2}
\end{minipage}
\caption{}
\end{figure}

For all ``staircase directions", in order that the Jost solutions be consistent with each of the equations \eqref{eq:nkevolution2}, we redefine their boundary conditions \eqref{eq:bcphis} \eqref{eq:bcpsis} as follows:
\begin{subequations}
\begin{align}\label{eq:bcphisI}
&\vp\sim\prod_{r\in\cI}\left(\frac{p_r+\zeta}{p_r-\zeta}\right)^{\frac12n_r}, \;\;\;\;\; \rvp\sim\prod_{r\in\cI}\left(\frac{p_r-\zeta}{p_r+\zeta}\right)^{\frac12n_r}  \;\;\;\;\; {\rm as} \;\; i\to-\infty \\
\label{eq:bcpsisI}
&\psi\sim\prod_{r\in\cI}\left(\frac{p_r-\zeta}{p_r+\zeta}\right)^{\frac12n_r}, \;\;\;\;\; \rpsi\sim\prod_{r\in\cI}\left(\frac{p_r+\zeta}{p_r-\zeta}\right)^{\frac12n_r} \;\;\;\;\; {\rm as} \;\: i\to+\infty.
\end{align}
\end{subequations}
When one restricts these boundary conditions to $\Gamma$ these agree with the previous boundary conditions \eqref{eq:bcphis} \eqref{eq:bcpsis}.  Now for each ``staircase direction", $\vp$ and $\rvp$ are linearly independent functions of equation \eqref{eq:nkevolution}, so the functions $\az$ and $\bz$, and the normalisation constants $\cz$, are independent of all ``staircase variables" $n_k$, $k\in\cI$. 

\subsection{``Non-Staircase Directions" ($\notin\cI$)} $\\$

We now consider how the spectral functions $\az$ and $\bz$ and the normalization constants $\cz$ evolve as we move in any of the ``non-staircase" lattice directions ($\notin\cI$). We denote the collection of these ``non-staircase" direction by $\cJ$, so that $\cI\cup\cJ$ is the entire N-dimensional lattice. For example in Figure \ref{fig:staircasetimeevo2} the two lattice directions in the lower plane are elements of $\cI$, but the orthogonal vertical direction is an element of $\cJ$. In order that the evolution of \eqref{eq:Q3} be well-defined in each of these directions we make the following assumption on the boundary conditions of $u$ at one end of the staircase:
\newline
\par
{\it In the limit $i\to-\infty$, the boundary conditions \eqref{eq:uBC-infty} for $u$, and \eqref{eq:cUBC-infty} for $\cU$, are assumed to remain unchanged under any shift in a ``non-staircase" direction ($\in\cJ$).}
\newline
\par
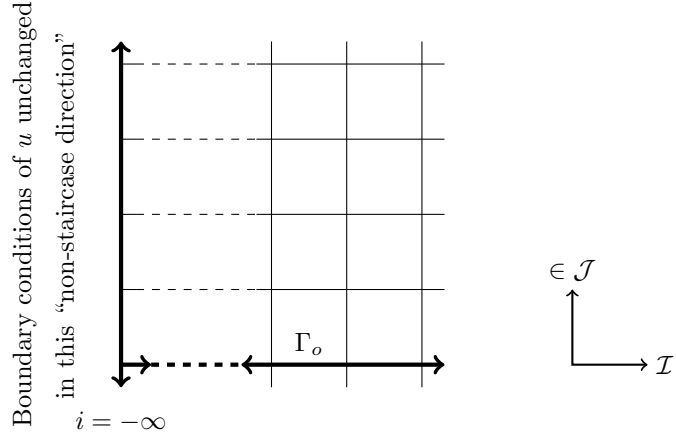
\begin{figure}[h!]
\begin{tikzpicture}[scale=1]
\draw [ultra thick, ->] (0,0) -- (0.4,0);
\draw [dashed, ultra thick] (0.4,0) -- (1.6,0);
\draw [ultra thick, <->] (1.6,0) -- (4.3,0);
\draw (0,1) -- (0.2,1);
\draw [dashed] (0.2,1) -- (1.8,1);
\draw (1.8,1) -- (4.3,1);
\draw (0,2) -- (0.2,2);
\draw [dashed] (0.2,2) -- (1.8,2);
\draw (1.8,2) -- (4.3,2);
\draw (0,3) -- (0.2,3);
\draw [dashed] (0.2,3) -- (1.8,3);
\draw (1.8,3) -- (4.3,3);
\draw (0,4) -- (0.2,4);
\draw [dashed] (0.2,4) -- (1.8,4);
\draw (1.8,4) -- (4.3,4);
\draw  [<->, ultra thick] (0,-0.3) -- (0,4.3);
\draw  (4,-0.3) -- (4,4.3);
\draw  (3,-0.3) -- (3,4.3);
\draw  (2,-0.3) -- (2,4.3);
\draw [thick, <->] (6,1) -- (6,0) -- (7,0);
\node [right] at (7,0) {$\cI$};
\node [above] at (6,1) {$\in\cJ$};
\node [below] at (0,-0.5) {$i=-\infty$};
\node [above] at (2.5,0) {$\Gamma_o$};
\node [left] at (-1,2) {\rotatebox{90}{Boundary conditions of $u$ unchanged}};
\node [left] at (-0.5,2) {\rotatebox{90}{in this ``non-staircase direction"}};
\end{tikzpicture}
\caption{Boundary conditions giving sufficient data to iterate the solution in the ``non-staircase" vertical direction}
\label{fig:staircasetimeevo4}
\end{figure}

This assuption gives sufficient data for the solution $u$ to be determined uniquely at all lattice points using \eqref{eq:Q3}. This is most evident when we consider a line of initial values $\Gamma_o$ as shown in Figure \ref{fig:staircasetimeevo4}. In this case $\cI$ is the horizontal direction, and $\cJ$ is the set of all such orthogonal vertical directions.

The assumption of the invariance of the boundary conditions of $u$ and $\cU$ (as $i\to-\infty$) gives sufficient data for the solution to be iterated to all points in the lattice. The boundary conditions \eqref{eq:bcphisI} \eqref{eq:bcpsisI} however are assumed to hold independently of these ``non-staircase variables" $n_s$. Since the evolution of the Jost solutions in any one ``non-staircase" direction is given by \eqref{eq:nkevolution}, our boundary conditions are inconsistent with \eqref{eq:nkevolution2}. To circumvent this we make the following definition.

\begin{definition}\label{def:jostN}
Let $\cJ$ denote the collection of ``non-staircase directions". The N-dimensional Jost solutions $\vp^{(N)},\rvp^{(N)}$ and $\psi^{(N)},\rpsi^{(N)}$, which are solutions to all N equations \eqref{eq:nkevolution}, are defined to be
\begin{subequations}
\begin{align}\label{eq:vpN}
&\vp^{(N)}:=\vp\prod_{s\in\cJ}\left(\frac{p_s+\zeta}{p_s-\zeta}\right)^{\frac12n_s}, \;\;\;\;\; \rvp^{(N)}:=\rvp\prod_{s\in\cJ}\left(\frac{p_s-\zeta}{p_s+\zeta}\right)^{\frac12n_s} \\
\label{eq:psiN}
&\psi^{(N)}:=\psi\prod_{s\in\cJ}\left(\frac{p_s-\zeta}{p_s+\zeta}\right)^{\frac12n_s}, \;\;\;\;\; \rpsi^{(N)}:=\rpsi\prod_{s\in\cJ}\left(\frac{p_s+\zeta}{p_s-\zeta}\right)^{\frac12n_s}.
\end{align}
\end{subequations}
\end{definition}

We then have the following result about the dependence of the spectral functions $\az$ and $\bz$ and the normalisation constants $\cz$ on each of the ``non-staircase variables" $n_s$, $s\in\cJ$. 

\begin{theorem} $\\$

\begin{enumerate}
\item[-] The function $\az$ is {\bf independent of all lattice variables}:
\begin{equation}
\az(n_s;\zeta)=\az(\zeta)
\end{equation}
\item[-] The dependence of the functions $\bz$ and $\cz$ on $n_s$, $s\in\cJ$, are given by
\begin{align}
\bz(n_s;\zeta)&=\bz(\zeta)\prod_{s\in\cJ}\left(\frac{p_s+\zeta}{p_s-\zeta}\right)^{n_s} \\
\cz(n_s;\zeta)&=\cz(\zeta)\prod_{s\in\cJ}\left(\frac{p_s+\zeta}{p_s-\zeta}\right)^{n_s}.
\end{align}
\end{enumerate}
\begin{proof}
Consider the ``non-staircase" $n_s$-direction, where $s\in\cJ$. The evolution equation for the N-dimensional Jost solutions in this direction is
\[
(p_s^2-\zeta^2)^{\frac12}\asf{\asf\vp}\xx^{(N)}-\left(\frac{P_s(\asf{\asf u}-u)}{\asf{\cU}}\right)\asf\vp\xx^{(N)}+(p_s^2-\zeta^2)^{\frac12}\vp^{(N)}=0,
\]
and by their boundary conditions the functions $\vp^{(N)}$ and $\rvp^{(N)}$ are linearly independent solutions of this equation. We may therefore write
\[
\psi^{(N)}=C_1\xx\rvp^{(N)}+C_2\xx\vp^{(N)},
\]
where $C_1$ and $C_2$ are independent of $n_s$, but may depend on all other lattice variables. This is equivalent to
\[
\psi=C_1\xx\rvp+C_2\xx\vp\prod_{s\in\cJ}\left(\frac{p_s+\zeta}{p_s-\zeta}\right)^{n_s},
\]
and by comparing this with 
\[
\psi=\az\xx\rvp+\bz\xx\vp\prod_{r\in\cI}\left(\frac{p_r+\zeta}{p_r-\zeta}\right)^{n_r},
\]
it follows that $\az$ is independent of $n_s$, while the dependence of $\bz$ on $n_s$ is through the plane-wave factor $\left(\frac{p_s+\zeta}{p_s-\zeta}\right)^{n_s}$. Repeating this argument for all $s\in\cJ$ gives the desired results for $\az$ and $\bz$. The result for $\cz$ follows from its definition \eqref{eq:nc}.
\end{proof}
\end{theorem}

\section{Inverse Problem for $\phi_1$}\label{sec:phi1inverse}

We now consider the inverse problem for $\phi_1$, that is the construction of the Jost solutions as functions of all N lattice variables. In doing so we alter the notation of all eigenfunctions by writing
\[
\vp(n_1,...,n_N;\zeta)\to\vp(\zeta),
\]
and similarly for the other functions. Here $\vp(\zeta)$ is understood to depend on {\it all} lattice variables $n_1,...,n_N$, however for the inverse problem it is convenient to suppress this dependence in the notation.
\par
Consider the equation \eqref{eq:ab}, which we rewrite as
\begin{equation}\label{eq:jump}
\frac{\up(\zeta)}{\az(\zeta)}-\rlam(\zeta)=R(\zeta)\lam(\zeta)\rho(\zeta),
\end{equation}
where the reflection coefficient $R$ is given by
\[
R(\zeta)=\frac{\bz(\zeta)}{\az(\zeta)}
\]
and the plane-wave factors $\rho$ are defined by
\[
\rho(\zeta):=\prod_{r\in\cI}\left(\frac{p_r+\zeta}{p_r-\zeta}\right)^{n_r}\prod_{s\in\cJ}\left(\frac{p_s+\zeta}{p_s-\zeta}\right)^{n_s}=\prod_{r=1}^N\left(\frac{p_r+\zeta}{p_r-\zeta}\right)^{n_r}.
\]
This defines a jump condition between two sectionally meromorphic functions along the contour ${\rm Re}(\zeta)=0$. The functions $\dfrac{\up}{\az}$ and $\rlam$ are analytic in the interior of the regions $\cR^+$ and $\cR^-$ respectively, and both are continuous along ${\rm Re}(\zeta)=0$. Given the jump condition and their boundary conditions, the question of determining a function which is equal to these in their respective half-planes is a Riemann-Hilbert problem. The method of solving such a problem is well-known (see e.g. \cite{g:66} Section 4): Consider the singular integral
\begin{align}\label{eq:singint}
&\hspace{0.5in}\frac1{2\pi i}\int_{-i\infty}^{+i\infty}\frac{R(\sigma)\lam(\sigma)}{\sigma+\zeta}\rho(\sigma)\xx d\sigma=\frac1{2\pi i}\int_{-i\infty}^{+i\infty}\frac{\up(\sigma)}{\az(\sigma)(\sigma+\zeta)}\xx d\sigma-\frac1{2\pi i}\int_{-i\infty}^{+i\infty}\frac{\rlam(\sigma)}{\sigma+\zeta}\xx d\sigma
\end{align}
where $\zeta\in\cR^+$. Here the contour of integration is the imaginary $\sigma$-axis. Since the function $\dfrac{\up}{\az}$ has $M$ simple poles in $\cR^+$ and has the boundary behaviour $\frac{\up}{\az}\sim1$ as $|\zeta|\to\infty$, one may use the residue theorem to calculate
\[
\frac1{2\pi i}\int_{-i\infty}^{+i\infty}\frac{\up(\sigma)}{\az(\sigma)(\sigma+\zeta)}\xx d\sigma=\frac12-\sum_{k=1}^M\frac{\up(\zeta_k)}{\az'(\zeta_k)(\zeta+\zeta_k)}.
\]
By then using the fact that
\[
\up(\zeta_k)=\bz(\zeta_k)\lam(\zeta_k)\rho(\zeta_k)
\]
this can written as
\[
\frac1{2\pi i}\int_{-i\infty}^{+i\infty}\frac{\up(\sigma)}{\az(\sigma)(\sigma+\zeta)}\xx d\sigma=\frac12-\sum_{k=1}^M\frac{\cz_k\lam(\zeta_k)}{(\zeta+\zeta_k)}\rho(\zeta_k).
\]
Secondly since $\rlam$ is analytic in $\cR^-$ one can determine that
\[
\frac1{2\pi i}\int_{-i\infty}^{+i\infty}\frac{\rlam(\sigma)}{\sigma+\zeta}\xx d\sigma=-\frac12+\rlam(-\zeta)=-\frac12+\lam(\zeta),
\]
and thus the singular integral equation becomes
\begin{equation}\label{eq:lamint}
\lam(\zeta)=1-\sum_{k=1}^M\frac{\cz_k\lam(\zeta_k)}{(\zeta+\zeta_k)}\rho(\zeta_k)-\frac1{2\pi i}\int_{-i\infty}^{+i\infty}\frac{R(\sigma)\lam(\sigma)}{\sigma+\zeta}\rho(\sigma)\xx d\sigma.
\end{equation}
Given the $\zeta_k$, the normalisations constants $\cz_k$ and the reflection coefficient $R$, one can use this equation to determine the Jost solution $\lam$ {\it as a function of $\zeta$ and all {\rN} lattice variables}. Remarkably all of the dependence on the lattice variables is contained in the plane-wave factors $\rho$. If instead we started with the relation
\begin{equation}
\lam(\zeta)=\az(\zeta)\rup(\zeta)-\bz(-\zeta)\up(\zeta)\rho(-\zeta),
\end{equation}
which is consistent with \eqref{eq:ab}, then by following the same procedure as above one finds that $\up(\zeta)$ is determined by solving the singular integral equation
\begin{equation}\label{eq:upint}
\up(\zeta)=1-\sum_{k=1}^M\frac{\dz_k\up(\zeta_k)}{(\zeta+\zeta_k)}\rho(-\zeta_k)+\frac1{2\pi i}\int_{-i\infty}^{+i\infty}\frac{S(\sigma)\up(\sigma)}{\sigma+\zeta}\rho(-\sigma)\xx d\sigma,
\end{equation}
where
\begin{equation}
\dz_k:=\frac1{\az'(\zeta_k)\bz(\zeta_k)}, \;\;\;\;\; S(\zeta):=\frac{\bz(-\zeta)}{\az(\zeta)}.
\end{equation}

\section{Inverse Problem for $\phi_2$}\label{sec:phi2inverse}

Now that we have constructed the first component of the eigenfunction $\bphi$ of \eqref{eq:Laxk}, we use the first component of the Lax equations to determine $\phi_2$: 
\begin{equation}\label{eq:phi2}
(\zeta^2-b^2)\xx\phi_2(\zeta)=(p_k^2-\zeta^2)^{\frac12}\cU\xx\asf{\phi}_1(\zeta)-(P_k\fru-(p_k^2-b^2)\xx u\xx)\xx\phi_1(\zeta).
\end{equation}
Let $\vp_2^{(N)}$ and $\psi_2^{(N)}$ be the corresponding second components for the Jost solutions $\vp^{(N)}$ and $\psi^{(N)}$ respectively, and define the functions $\lam_2$ and $\up_2$ by
\[
\vp_2^{(N)}=\lam_2\prod_{r=1}^N\left(\frac{p_r+\zeta}{p_r-\zeta}\right)^{n_r}, \;\;\;\;\; \psi_2^{(N)}=\up_2\prod_{r=1}^N\left(\frac{p_r-\zeta}{p_r+\zeta}\right)^{n_r}.
\]
Then $\lam_2$ is given by 
\begin{equation}\label{eq:lam2}
(\zeta^2-b^2)\xx\lam_2(\zeta)=(p_k+\zeta)\xx\cU\xx\asf{\lam}(\zeta)-(P_k\fru-(p_k^2-b^2)\xx u\xx)\xx\lam(\zeta),
\end{equation}
and $\up_2$ is given by
\begin{equation}\label{eq:up2}
(\zeta^2-b^2)\xx\up_2(\zeta)=(p_k-\zeta)\xx\cU\xx\asf{\up}(\zeta)-(P_k\fru-(p_k^2-b^2)\xx u\xx)\xx\up(\zeta).
\end{equation}
We see that $\lam_2(\zeta)$ and $\up_2(\zeta)$ are analytic in $\cR^+$, except for a simple pole at $\zeta=b$. Thus $\rlam_2(\zeta)=\lam_2(-\zeta)$ and $\rup_2(\zeta)=\up_2(-\zeta)$, which are the second components for the Jost solutions $\rlam$ and $\rup$ respectively, are analytic in $\cR^-$ except for a simple pole at $\zeta=-b$. Furthermore all of these functions are continuous on the imaginary $\zeta$-axis. Now since the two eigenfunctions 
\[
{\boldsymbol\vp}^{(N)}:=\left(\begin{array}{c} \vp^{(N)}  \\ \vp_2^{(N)} \end{array}\right), \;\;\;\;\; {\boldsymbol\rvp}^{(N)}:=\left(\begin{array}{c} \rvp^{(N)}  \\ \rvp_2^{(N)}  \end{array}\right),
\]
are linearly independent solutions of the Lax equations \eqref{eq:Laxk}, we may write
\[
{\boldsymbol\psi}^{(N)} :=\left(\begin{array}{c} \psi^{(N)}  \\ \psi_2^{(N)}  \end{array}\right)=\az(\zeta)\xx{\boldsymbol\rvp}^{(N)}+\bz(\zeta)\xx{\boldsymbol\vp}^{(N)},
\]
whose second component may be written in terms of $\lam_2$ and $\up_2$ as
\begin{equation}\label{eq:jump2}
\frac{\up_2(\zeta)}{\az(\zeta)}-\rlam_2(\zeta)=R(\zeta)\xx\lam_2(\zeta)\xx\rho(\zeta).
\end{equation}
Note that these are the same functions $\az$ and $\bz$ as those that appear in the integral equation \eqref{eq:lamint} for $\lam(\zeta)$. As in the inverse problem for $\phi_1$, equation \eqref{eq:jump2} becomes the jump condition for a Riemann-Hilbert problem, and as such we look at the singular integral
\begin{align}\label{eq:lam2int1}
&\hspace{0.5in}\frac1{2\pi i}\int_{-i\infty}^{+i\infty}\frac{R(\sigma)\lam_2(\sigma)}{\sigma+\zeta}\rho(\sigma)\xx d\sigma=\frac1{2\pi i}\int_{-i\infty}^{+i\infty}\frac{\up_2(\sigma)}{\az(\sigma)(\sigma+\zeta)}\xx d\sigma-\frac1{2\pi i}\int_{-i\infty}^{+i\infty}\frac{\rlam_2(\sigma)}{\sigma+\zeta}\xx d\sigma.
\end{align}
The difference in this case is that $\up_2$ and $\rlam_2$ have simple poles at $+b$ and $-b$ respectively, and both of these function are $\cO(\xx\zeta^{-1}\xx)$ as $|\zeta|\to\infty$. Let us first consider the integral involving $\up_2$. By using the residue theorem one has
\begin{equation}\label{eq:up2int1}
\frac1{2\pi i}\int_{-i\infty}^{+i\infty}\frac{\up_2(\sigma)}{\az(\sigma)(\sigma+\zeta)}\xx d\sigma=-\frac{{\rm Res}_{\zeta=b}\bigl[\up_2\bigr]}{\az(b)(\zeta+b)}-\sum_{k=1}^M\frac{\cz_k\lam_2(\zeta_k)}{(\zeta+\zeta_k)}\rho(\zeta_k).
\end{equation}
From the Lax equations however, by eliminating the first component $\phi_1$ one can show that at $\zeta=b$ the second-order linear equation for $\phi_2$ in the $n_k$-direction drastically simplifies to
\begin{equation}\label{eq:phi2atb}
(p_k^2-a^2)^{\frac12}\xx\asf{\asf\phi}_2-\left(\frac{P_k(\asf{\asf u}-u)}{\asf{\cU}}\right)\asf\phi_2+(p_k^2-a^2)^{\frac12}\xx\phi_2=0,
\end{equation}
which we identify as equation \eqref{eq:nkevolution} at $\zeta=a$. Since this holds for every lattice direction we may write
\[
\bigl.(\zeta-b)\psi_2^{(N)}(\zeta)\bigr|_{\zeta=b}=\alpha\xx\vp^{(N)}(a)+\beta\xx\psi^{(N)}(a)
\]
for some constants $\alpha$ and $\beta$. In terms of $\up_2$ and $\lam_2$ this implies that
\begin{align}
\bigl.(\zeta-b)\up_2(\zeta)\bigr|_{\zeta=b}=&\alpha\xx\lam(a)\prod_{r=1}^N\left(\frac{(p_r+a)(p_r+b)}{(p_r-a)(p_r-b)}\right)^{\frac12n_r}+\beta\xx\up(a)\prod_{r=1}^N\left(\frac{(p_r-a)(p_r+b)}{(p_r+a)(p_r-b)}\right)^{\frac12n_r} \nonumber \\
\label{eq:up2atb}
=&\alpha\xx\lam(a)\xx\cF(a,b)+\beta\xx\up(a)\xx\cF(-a,b),
\end{align}
where the plane-wave factors $\cF$ are defined by
\[
\cF(a,b):=\prod_{r=1}^N\left(\frac{(p_r+a)(p_r+b)}{(p_r-a)(p_r-b)}\right)^{\frac12n_r}.
\]
In order to determine $\alpha$ and $\beta$ we consider equation \eqref{eq:up2atb} in the limit $i\to\pm\infty$. Firstly we have the asymptotic behaviour
\begin{align*}
&\lam(a)\sim\az(a)+c_o\rho(-a) \;\; {\rm as} \;\; i\to+\infty \\
&\up(a)\sim\az(a)+c_1\rho(a) \;\; {\rm as} \;\; i\to-\infty,
\end{align*}
for some constants $c_o$ and $c_1$. Since $a>0$ both of these plane-wave factors are exponentially small. Thus using the boundary conditions of $u$ and $\cU$ in \eqref{eq:up2}, as $i\to-\infty$ we have
\begin{align*}
\bigl.(\zeta-b)\up_2(\zeta)\bigr|_{\zeta=b}&\sim\frac{\az(b)}{2b}\Bigl[(p_k-b)\xx\cU\xx-(P_k\fru-(p_k^2-b^2)\xx u\xx)\Bigr] \\
&\sim(a-b)\az(b)\cC\cF(-a,b),
\end{align*}
which implies that $\beta\az(a)=(a-b)\xx\cC\az(b)$. By then taking $i\to+\infty$ we have
\begin{align*}
\bigl.(\zeta-b)\up_2(\zeta)\bigr|_{\zeta=b}&\sim\frac1{2b}\Bigl[(p_k-b)\xx\cU\xx-(P_k\fru-(p_k^2-b^2)\xx u\xx)\Bigr] \\
&\sim-(a+b)\xx\cK_o\cA\cF(a,b),
\end{align*}
which gives $\alpha\az(a)=-(a+b)\xx\cK_o\cA$. Therefore the integral in equation \eqref{eq:up2int1} becomes
\begin{align*}
\frac1{2\pi i}\int_{-i\infty}^{+i\infty}\frac{\up_2(\sigma)}{\az(\sigma)(\sigma+\zeta)}\xx d\sigma=&\;(a+b)\left(\frac{\cK_o}{\az(a)\az(b)}\right)\cA\cF(a,b)\left(\frac{\lam(a)}{\zeta+b}\right)-(a-b)\cC\cF(-a,b)\left(\frac{\up(a)}{\az(a)(\zeta+b)}\right) \\
&-\sum_{k=1}^M\frac{\cz_k\lam_2(\zeta_k)}{(\zeta+\zeta_k)}\rho(\zeta_k).
\end{align*}
The integral involving $\rlam$ in \eqref{eq:lam2int1} can be evaluated to be
\begin{equation}\label{eq:rlam2int}
\frac1{2\pi i}\int_{-i\infty}^{+i\infty}\frac{\rlam_2(\sigma)}{\sigma+\zeta}\xx d\sigma=\lam_2(\zeta)+\frac1{\zeta-b}\lim_{\sigma\to -b}\Bigl[\rlam_2(\sigma)(\sigma+b)\Bigr],
\end{equation}
and by similar reasoning and using the fact that $\rlam(-a)=\lam(a)$ etc., we find
\begin{align*}
\lim_{\sigma\to -b}\Bigl[\rlam_2(\sigma)(\sigma+b)\Bigr]=&\;-(a+b)\left(\frac{\cK_1\az(b)}{\az(a)}\right)\cB\cF(a,-b)\lam(a)+(a+b)\cD\cF(-a,-b)\left(\frac{\up(a)}{\az(a)}\right).
\end{align*}
Thus the integral \eqref{eq:rlam2int} becomes
\begin{align*}
\frac1{2\pi i}\int_{-i\infty}^{+i\infty}\frac{\rlam_2(\sigma)}{\sigma+\zeta}\xx d\sigma=\lam_2(\zeta)-(a-b)\left(\frac{\cK_1\az(b)}{\az(a)}\right)\cB\cF(a,-b)\left(\frac{\lam(a)}{\zeta-b}\right)+(a+b)\cD\cF(-a,-b)\left(\frac{\up(a)}{\az(a)(\zeta-b)}\right).
\end{align*}
Using these in \eqref{eq:lam2int1} then gives the following closed-form singular integral equation for $\lam_2$:
\begin{align}
\lam_2(\zeta)&=(a+b)\left(\frac{\cK_o}{\az(a)\az(b)}\right)\cA\cF(a,b)\left(\frac{\lam(a)}{\zeta+b}\right)+(a-b)\left(\frac{\cK_1\az(b)}{\az(a)}\right)\cB\cF(a,-b)\left(\frac{\lam(a)}{\zeta-b}\right) \\
&\;-(a-b)\cC\cF(-a,b)\left(\frac{\up(a)}{\az(a)(\zeta+b)}\right)-(a+b)\cD\cF(-a,-b)\left(\frac{\up(a)}{\az(a)(\zeta-b)}\right) \\
\label{eq:lam2int}
&-\sum_{k=1}^M\left(\frac{\cz_k\lam_2(\zeta_k)}{(\zeta+\zeta_k)}\right)\xx\rho(\zeta_k)-\frac1{2\pi i}\int_{-i\infty}^{+i\infty}\left(\frac{R(\sigma)\lam_2(\sigma)}{\sigma+\zeta}\right)\xx\rho(\sigma)\xx d\sigma.
\end{align}
A natural ansatz for this is
\begin{align}
\lam_2(\zeta)&=(a+b)\left(\frac{\cK_o}{\az(a)\az(b)}\right)\cA\cF(a,b)\lam(a)\xi_b(\zeta)+(a-b)\left(\frac{\cK_1\az(b)}{\az(a)}\right)\cB\cF(a,-b)\lam(a)\xi_{-b}(\zeta) \nonumber \\
\label{eq:lam2ansatz}
&\;-(a-b)\cC\cF(-a,b)\left(\frac{\up(a)}{\az(a)}\right)\xi_b(\zeta)-(a+b)\cD\cF(-a,-b)\left(\frac{\up(a)}{\az(a)}\right)\xi_{-b}(\zeta)
\end{align}
where by equation \eqref{eq:lam2int} the functions $\xi_{\pm b}(\zeta)$ are calculated by solving
\begin{equation}\label{eq:xpmb}
\xi_{\pm b}(\zeta)=\frac1{\zeta\pm b}-\sum_{k=1}^M\left(\frac{\cz_k\xx\xi_{\pm b}(\zeta_k)}{\zeta+\zeta_k}\right)\rho(\zeta_k)-\frac1{2\pi i}\int_{-i\infty}^{+i\infty}\left(\frac{R(\sigma)\xx\xi_{\pm b}(\sigma)}{\sigma+\zeta}\right)\rho(\sigma)\xx d\sigma.
\end{equation}
Note that the ingredients in these equations are the scattering data from the forward scattering of $\phi_1$, and \eqref{eq:xpmb} differs from \eqref{eq:lamint} only in the source term.

\begin{remark}
If we define the quantity
\begin{equation}\label{eq:nqcvar}
S(a,b):=\frac1{a+b}-\xi_b(a)
\end{equation}
then by comparing \eqref{eq:xpmb} with the integral equation considered in the direct linearization approach \cite{nqc:83} we see that this object is in fact a solution (containing solitons and radiation) of the NQC equation.
\end{remark}

\section{Reconstruction of the Solution of Q3$_\delta$}\label{sec:ureconstruction}

We now show how one can recontruct the solution $u$ as a function of all N lattice variables. Consider equation \eqref{eq:lam2}, which holds for any $n_k,\;k=1,...,N$. Firstly by dividing through by $(p_k+\zeta)$ and taking the limit $|\zeta|\to\infty$ we have
\begin{equation}\label{eq:Usol}
\cU=\lim_{|\zeta|\to\infty}\bigl[\xx\zeta\lam_2(\zeta)\xx\bigr].
\end{equation}
Using equations \eqref{eq:lam2ansatz} and \eqref{eq:xpmb} this may be expressed as
\begin{align}
\cU&=(a+b)\left(\frac{\cK_o}{\az(a)\az(b)}\right)\cA\cF(a,b)\lam(a)V(b)+(a-b)\left(\frac{\cK_1\az(b)}{\az(a)}\right)\cB\cF(a,-b)\lam(a)V(-b) \nonumber \\
\label{eq:Urepresentation}
&\;-(a-b)\cC\cF(-a,b)\left(\frac{\up(a)}{\az(a)}\right)V(b)-(a+b)\cD\cF(-a,-b)\left(\frac{\up(a)}{\az(a)}\right)V(-b), \\
\end{align}
where $V(\pm b)$ is given by
\begin{equation}
V(\pm b)=1-\sum_{k=1}^M\cz_k\xi_{\pm b}(\zeta_k)\rho(\zeta_k)-\frac1{2\pi i}\int_{-i\infty}^{+i\infty}R(\sigma)\xi_{\pm b}(\sigma)\rho(\sigma)\xx d\sigma.
\end{equation}
In fact it turns out that $\lam$ and $V$ are the same object. Equation \eqref{eq:lam2} could then in principle be summed to find $u$. There is a way however to obtain a closed-form expression for $u$ rather than its derivative. To do this we first note that in the Lax equations \eqref{eq:Laxk}, one is free to interchange the roles of the parameters $a$ and $b$. In other words the N equations
\begin{equation}\label{eq:Laxka}
(p_k^2-\zeta^2)^{\frac12}\asf{\bphi}=\frac1{\cU}\left(\begin{array}{cc} P_k\fru-(p_k^2-a^2)u & \zeta^2-a^2 \vspace{0.1 in}\\ \cU\frU-\frac{\delta^2(p_k^2-a^2)}{4P_k(\zeta^2-a^2)} & (p_k^2-a^2)\fru-P_ku \end{array}\right)\bphi
\end{equation}
are also N Lax equations for Q3$_\delta$. Note however that we are {\it not} sqapping $a$ and $b$ in the functions $u$ and $\cU$, nor in the initial conditions, nor in the boundary conditions. We are simply repeating the entire IST with the new Lax equations \eqref{eq:Laxka} in place of \eqref{eq:Laxk}, which is permissible due to the symmetric dependence of Q3$_\delta$ on $a$ and $b$. There are however some important remarks to be made. Firstly the forward scattering problem \eqref{eq:ds11} for $\phi_1$ is independent of $a$ and $b$, and thus {\it interchanging $a$ and $b$ in the Lax equations will not change the Jost solutions $\lam$ and $\up$, nor the scattering data}. The second component $\phi_2$ however will now be calculated by swapping $a$ and $b$ in the Lax equations \eqref{eq:lam2} and \eqref{eq:up2}, and re-deriving the integral equation \eqref{eq:xpmb}. To make a clear distinction between the original quantities $\lam_2, \up_2$ and their new counterparts, we rewrite the original functions as
\[
\lam_2\to\lam_2^{(b)}, \;\;\;\;\; \up_2\to\up_2^{(b)},
\]
and then denote the new functions, which are  obtained by swapping $a$ and $b$ in \eqref{eq:lam2} and \eqref{eq:up2}, by $\lam_2^{(a)}$ and $\up_2^{(a)}$. By repeating the analysis of the previous section we find
\begin{align}
\lam_2^{(a)}(\zeta)&=(a+b)\left(\frac{\cK_o}{\az(a)\az(b)}\right)\cA\cF(a,b)\lam(b)\xi_a(\zeta)+(a-b)\left(\frac{\cK_1\az(b)}{\az(a)}\right)\cB\cF(a,-b)\left(\frac{\up(b)}{\az(b)}\right)\xi_{a}(\zeta) \nonumber \\
\label{eq:lam2aansatz}
&\;-(a-b)\cC\cF(-a,b)\lam(b)\xi_{-a}(\zeta)-(a+b)\cD\cF(-a,-b)\left(\frac{\up(b)}{\az(b)}\right)\xi_{-a}(\zeta).
\end{align}
Then finally given the quantities $\lam_2^{(b)}$ and $\lam_2^{(a)}$ one can combine the two versions of \eqref{eq:lam2}, namely
\begin{align*}
(\zeta^2-b^2)\xx\lam_2^{(b)}(\zeta)&=(p_k+\zeta)\xx\cU\xx\asf{\lam}(\zeta)-(P_k\fru-(p_k^2-b^2)\xx u\xx)\xx\lam(\zeta) \\
(\zeta^2-a^2)\xx\lam_2^{(a)}(\zeta)&=(p_k+\zeta)\xx\cU\xx\asf{\lam}(\zeta)-(P_k\fru-(p_k^2-a^2)\xx u\xx)\xx\lam(\zeta) 
\end{align*}
to express the N-dimensional solution of Q3$_\delta$ as
\begin{equation}\label{eq:usol1}
u=\frac{(\zeta^2-b^2)\xx\lam_2^{(b)}(\zeta)-(\zeta^2-a^2)\xx\lam_2^{(a)}(\zeta)}{(a^2-b^2)\xx\lam(\zeta)}.
\end{equation}
Since however the solution is independent of $\zeta$, we may take it to be large, in which case the solution can be expressed as
\begin{equation}\label{eq:usol}
u=\lim_{|\zeta|\to\infty}\left(\xx\frac{\zeta^2\bigl[\lam_2^{(b)}(\zeta)-\lam_2^{(a)}(\zeta)\bigr]}{a^2-b^2}\xx\right).
\end{equation}
This solution is obtained by solving the integral equations \eqref{eq:lamint}, \eqref{eq:upint} and \eqref{eq:xpmb} for $\lam, \up$ and $\xi$, using these objects in the expressions \eqref{eq:lam2ansatz} and \eqref{eq:lam2aansatz} for $\lam^{(b)}$ and $\lam^{(a)}$ respectively, and then taking the limit in \eqref{eq:usol}.

\begin{remark}$\\$

For the reflectionless case $\bz\equiv0$ the Jost solutions exist at the points $\zeta=-a$ and $\zeta=-b$, and we have $\cK_o=\az(a)\az(b)$ and $\cK_1=\az(a)/\az(b)$. Comparing the expressions for $u$ obtained by setting $\zeta=\pm a, \pm b$ in \eqref{eq:usol1}, and using the fact that $\up(\zeta)=\az(\zeta)\lam(-\zeta)$ we find
\begin{equation}\label{eq:ureflectionless}
u=(a+b)\cA\cF(a,b)\xi_{a}(b)+(a-b)\cB\cF(a,-b)\xi_{a}(-b)-(a-b)\cC\cF(-a,b)\xi_{-a}(b)-(a+b)\cD\cF(-a,-b)\xi_{-a}(-b).
\end{equation}
Using the NQC variable $S(a,b)$ defined by \eqref{eq:nqcvar} this becomes
\begin{align}
u=&\xx\cA\cF(a,b)\bigl[1-(a+b)S(a,b)\bigr]+\cB\cF(a,-b)\bigl[1-(a-b)S(a,-b)\bigr] \nonumber \\
\label{eq:Q3NsolIS}
+&\xx\cC\cF(-a,b)\bigl[1+(a-b)S(-a,b)\bigr]+\cD\cF(-a,-b)\bigl[1+(a+b)S(-a,-b)\bigr].
\end{align}
This is precisely the form of the N-soliton solution of Q3$_\delta$ obtained in \cite{nah:09}. The corresponding dual function $\cU$ is given by
\begin{align}
\cU=&\xx(a+b)\cA\cF(a,b)\lam(a)\lam(b)+(a-b)\cB\cF(a,-b)\lam(a)\lam(-b) \nonumber \\
\label{eq:Q3NsolUIS}
-&\xx(a-b)\cC\cF(-a,b)\lam(-a)\lam(b)-(a+b)\cD\cF(-a,-b)\lam(-a)\lam(-b),
\end{align}
which was also obtained in \cite{nah:09}.
\end{remark}

\section{One-soliton Example}\label{sec:onesoliton}

Here we construct an explicit solution to Q3$_\delta$ for the special case where the reflection coefficient $R$ is identically zero and the function $\az$ has exactly one zero in $\cR^+$ at $\zeta_1=k>0$. The constants $\cK_o$ and $\cK_1$ are set to $\left(\frac{(a-k)(b-k)}{(a+k)(b+k)}\right)$ and $\left(\frac{(a-k)(b+k)}{(a+k)(b-k)}\right)$ respectively. We are required to solve the singular integral equation \eqref{eq:xpmb} for $\xi_{\pm b}$. We write the normalisation constant as $\cz_1=2kc$ with $c$ being a constant, which we absorb into the plane-wave factor $\rho$. Equation \eqref{eq:xpmb} becomes
\begin{equation}
\xi_{\pm b}(\zeta)=\frac1{\zeta\pm b}-\left(\frac{2k}{\zeta+k}\right)\xi_{\pm b}(\zeta_k)\rho(\zeta_k),
\end{equation}
which has the solution
\begin{equation}
\xi_{\pm b}(\zeta)=\frac1{\zeta\pm b}\left(\frac{1+\frac{(\zeta-k)(b\mp k)}{(\zeta+k)(b\pm k)}\rho(k)}{1+\rho(k)}\right).
\end{equation}
From \eqref{eq:ureflectionless} the solution is then 
\begin{align}
u&=\xx\cA\cF(a,b)\left(\frac{1+\left(\frac{(a-k)(b-k)}{(a+k)(b+k)}\right)\rho(k)}{1+\rho(k)}\right)+\cB\xx\cF(a,-b)\left(\frac{1+\left(\frac{(a-k)(b+k)}{(a+k)(b-k)}\right)\rho(k)}{1+\rho(k)}\right) \nonumber \\
&+\cC\xx\cF(-a,b)\left(\frac{1+\left(\frac{(a+k)(b-k)}{(a-k)(b+k)}\right)\rho(k)}{1+\rho(k)}\right)+\cD\xx\cF(-a,-b)\left(\frac{1+\left(\frac{(a+k)(b+k)}{(a-k)(b-k)}\right)\rho(k)}{1+\rho(k)}\right)
\end{align}
where $\cA\cD(a+b)^2-\cB\cC(a-b)^2=-\frac{\delta^2}{16ab}$.

\section{Conclusion}

In this chapter we have rigorously derived a discrete IST for the Q3$_\delta$ lattice equation. The initial-value space was given on a multidimensional staircase within an N-dimensional lattice, and we have given examples of how such staircases may be constructed. By incorporating the multidimensional consistency of the Q3$_\delta$ equation into the IST scheme we have shown how to solve the inverse problem for this equation for solutions depending on an N discrete independent variables. The assumptions made on the solution were that it be real and that the initial profile satisfy the summability condition \eqref{eq:potsum}, which is a much weaker condition than that imposed in \cite{b:12}. The solution to Q3$_\delta$ incorporating both solitons and radiation is found by solving the singular integral equations \eqref{eq:lamint}, \eqref{eq:upint} and \eqref{eq:xpmb}, and then taking the limit given in \eqref{eq:usol}. This method of solution should of course also apply to the other ABS equations, and will form the basis of future research.

\section{Appendix}

\subsection{Proof of Theorem \ref{th:jostanal1}}\label{sec:app1}

We first prove \eqref{eq:lamest1}. For $\zeta\in\cR^+,\;\zeta\neq0$ the recursion relation \eqref{eq:Hkrr} can be upper-bounded by
\begin{equation}\label{eq:Hkest1}
|H_{k+1}(i;\zeta)|\leq\sum_{l=-\infty}^{i-1}|\pot(l)||H_k(l;\zeta)|.
\end{equation}
We then claim that
\begin{equation}\label{eq:Hkest2}
|H_k(i;\zeta)|\leq\frac{F(i)^k}{k!},
\end{equation}
where
\[
F(i)=\sum_{r=-\infty}^{i-1}|\pot(r)|.
\]
Clearly this holds for $k=0$. To prove the inductive step we use \eqref{eq:Hkest1}, summation by parts and the fact that $F(i+1)\geq F(i)$: 
\begin{align*}
|H_{k+1}(i;\zeta)|&\leq\sum_{l=-\infty}^{i-1}|\pot(l)|\frac{F(l)^k}{k!} \\
&=\frac1{k!}\sum_{l=-\infty}^{i-1}\bigl[F(l+1)-F(l)\bigr]F(l)^k \\
&=\frac{F(i)^{k+1}}{k!}-\frac1{k!}\sum_{l=-\infty}^{i-1}\bigl[F(l+1)^k-F(l)^k\bigr]F(l+1) \\
&=\frac{F(i)^{k+1}}{k!}-\frac1{k!}\sum_{l=-\infty}^{i-1}\bigl[F(l+1)-F(l)\bigr]F(l+1)\left(\;\sum_{r=0}^{k-1}F(l+1)^{k-1-r}F(l)^r\;\right) \\
&\leq\frac{F(i)^{k+1}}{k!}-\frac k{k!}\sum_{l=-\infty}^{i-1}\bigl[F(l+1)-F(l)\bigr]F(l)^k,
\end{align*}
and so by examining the second and last lines we have
\[
\frac1{k!}\sum_{l=-\infty}^{i-1}\bigl[F(l+1)-F(l)\bigr]F(l)^k\leq\frac{F(i)^{k+1}}{(k+1)!},
\]
which then shows that
\[
|H_{k+1}(i;\zeta)|\leq\frac{F(i)^{k+1}}{(k+1)!}.
\]
Thus the estimate \eqref{eq:Hkest2} holds. The series solution for $\lam$ can then be upper-bounded by
\begin{align*}
|\lam(i;\zeta)-1|&\leq\sum_{k=1}^{+\infty}\frac{|H_k(i;\zeta)|}{|\zeta|^k}\leq\sum_{k=1}^{+\infty}\frac{F(i)^k}{|\zeta|^kk!}\leq\left(\frac{F(+\infty)}{|\zeta|}\right)\exp\left[\frac{F(+\infty)}{|\zeta|}\right]\leq C_1
\end{align*}
since \eqref{eq:potsum} holds. Thus \eqref{eq:lamest1} is proved and so for any $\zeta\neq0$ the series solution for $\lam$ converges absolutely and uniformly in $i$. We now prove \eqref{eq:lamest10}. To allow for $\zeta=0$ we give an alternative upper-bound for the summation equation \eqref{eq:lamsum}. One can easily verify that
\begin{align}\label{eq:exprewrite}
1-&\xx\prod_{r=l}^{i-1}\left(\frac{\pz(r)-\zeta}{\pz(r)+\zeta}\right)=\sum_{j=l}^{i-1}\left[1-\left(\frac{\pz(j)-\zeta}{\pz(j)+\zeta}\right)\right]\prod_{r=l}^{j-1}\left(\frac{\pz(r)-\zeta}{\pz(r)+\zeta}\right) \\
\Rightarrow |\lam(i;\zeta)|&\leq1+\sum_{l=-\infty}^{i-1}\left(\sum_{j=l}^{i-1}\frac1{|\pz(j)+\zeta|}\prod_{r=l}^{j-1}\left|\frac{\pz(r)-\zeta}{\pz(r)+\zeta}\right|\right)|\pot(l)||\lam(l;\zeta)| \nonumber \\
&\leq1+\sum_{l=-\infty}^{i-1}\left(\sum_{j=l}^{i-1}\frac1{|\pz(j)+\zeta|}\right)|\pot(l)||\lam(l;\zeta)|. \nonumber
\end{align}
For $\zeta\in\cR^+$ however we have
\[
|\pz(j)+\zeta|\geq|\pz(j)-\zeta|\geq\bigl|\xx|\pz(j)|-|\zeta|\xx\bigr|\geq|\pz(j)|-|\zeta|\geq|\pz(j)|
\]
and so we have
\begin{equation}\label{eq:lammajor1}
|\lam(i;\zeta)|\leq1+\eta\sum_{l=-\infty}^{i-1}(i-l)\xx|\pot(l)|\xx|\lam(l;\zeta)|,
\end{equation}
where $\eta=\max\{\xx|\pz(r)^{-1}| : r\in\cI\xx\}$. Equation \eqref{eq:lammajor1} is a majorant for both the summation equations \eqref{eq:lamsum} and \eqref{eq:lamsum0}, and thus may be used to estimate $\lam(i;\zeta)$ for all $\zeta\in\cR^+$. Thus we have
\[
|\lam(i;\zeta)|\leq\sum_{k=0}^{+\infty}\eta^kH_k^*(i)
\]
where
\[
H_0^*=1, \;\;\;\;\; H_{k+1}^*(i)=\sum_{l=-\infty}^{i-1}(i-l)\xx|\pot(l)|\xx H_k^*(l).
\]
We claim that
\begin{equation}\label{eq:Hkoest2}
|H_k^*(i)|\leq\frac{G(i,i)^k}{k!}
\end{equation}
where
\[
G(i,j)=\sum_{r=-\infty}^{j-1}(i-r)\xx|\pot(r)|.
\]
Clearly this holds for $k=0$. To prove the inductive step we use the recursion relation for $H_k^*$, and again summation by parts and the properties of $G$:
\begin{align*}
|H_{k+1}^*(i)|&\leq\sum_{l=-\infty}^{i-1}(i-l)\xx|\pot(l)|\xx\left(\frac{G(l,l)^k}{k!}\right) \\
&\leq\frac1{k!}\sum_{l=-\infty}^{i-1}\bigl[G(i,l+1)-G(i,l)\bigr]G(i,l)^k \\
&=\frac1{k!}G(i,i)^{k+1}-\frac1{k!}\sum_{l=-\infty}^{i-1}\bigl[G(i,l+1)^k-G(i,l)^k\bigr]G(i,l+1) \\
&=\frac1{k!}G(i,i)^{k+1}-\frac1{k!}\sum_{l=-\infty}^{i-1}\bigl[G(i,l+1)-G(i,l)\bigr]G(i,l+1)\left(\;\sum_{r=0}^{k-1}G(i,l+1)^{k-1-r}G(i,l)^r\;\right) \\
&\leq\frac1{k!}G(i,i)^{k+1}-\frac{k}{k!}\sum_{l=-\infty}^{i-1}\bigl[G(i,l+1)-G(i,l)\bigr]G(i,l)^k,
\end{align*}
and again by examining the second and last lines we have
\[
\frac1{k!}\sum_{l=-\infty}^{i-1}\bigl[G(i,l+1)-G(i,l)\bigr]G(i,l)^k\leq\frac{G(i,i)^{k+1}}{(k+1)!}
\]
which implies
\[
|H_{k+1}^*(i)|\leq\frac{G(i,i)^{k+1}}{(k+1)!}
\]
and completes the inductive step. Thus for all $\zeta\in\cR^+$
\[
|\lam(i;\zeta)-1|\leq\sum_{k=1}^{+\infty}\frac{\eta^kG(i,i)^k}{k!}\leq\eta\xx G(i,i)\exp\xx\Bigl[\xx\eta\xx G(i,i)\Bigr].
\]
Let us first consider $i\leq0$. In this case we have
\begin{align*}
|\lam(i;\zeta)-1|&\leq\eta\exp\Bigl[\xx\eta\xx G(i,i)\Bigr]\left[\xx i\sum_{r=-\infty}^{i-1}|\pot(r)|+\sum_{r=-\infty}^{i-1}(-r)|\pot(r)|\xx\right] \\
&\leq\eta\exp\Bigl[\xx\eta\xx G(0,0)\Bigr]\left[\xx\sum_{r=-\infty}^{-1}(-r)|\pot(r)|\xx\right] \\
&\leq D_1,
\end{align*}
for some constant $D_1$, courtesy of \eqref{eq:potsum}. To examine the case $i>0$ we consider the majorant \eqref{eq:lammajor1}
\begin{align*}
|\lam(i;\zeta)|&\leq1+\eta\sum_{l=-\infty}^{i-1}(-l)\xx|\pot(l)|\xx|\lam(l;\zeta)|+i\xx\eta\sum_{l=-\infty}^{i-1}\xx|\pot(l)|\xx|\lam(l;\zeta)| \\
&\leq1+\eta\sum_{l=-\infty}^{-1}(-l)\xx|\pot(l)|\xx|\lam(l;\zeta)|+i\xx\eta\sum_{l=-\infty}^{i-1}\xx|\pot(l)|\xx|\lam(l;\zeta)| \\
&\leq D_2+i\xx\eta\sum_{l=-\infty}^{i-1}\xx|\pot(l)|\xx|\lam(l;\zeta)|
\end{align*}
for some constant $D_2$, where we have used \eqref{eq:potsum} and the fact that $\lam(i;\zeta)$ can be upper-bounded by a constant for $i\leq0$. Write $\lam(i;\zeta)=D_2(1+i)\Xi(i;\zeta)$, then the upper-bound for $\Xi$ becomes
\begin{align*}
|\Xi(i;\zeta)|&\leq1+\eta\left(\frac i{1+i}\right)\sum_{l=-\infty}^{i-1}(1+|l|)\xx|\pot(l)|\xx|\Xi(l;\zeta)| \\
&\leq1+\eta\sum_{l=-\infty}^{i-1}(1+|l|)\xx|\pot(l)|\xx|\Xi(l;\zeta)|.
\end{align*}
Making similar arguments to those presented above it follows that
\begin{align*}
|\Xi(i;\zeta)|&\leq\exp\left(\eta\sum_{l=-\infty}^{i-1}(1+|l|)\xx|\pot(l)|\right)\leq\exp\left(\eta\sum_{l=-\infty}^{+\infty}(1+|l|)\xx|\pot(l)|\right)
\end{align*}
which in turn implies that for $i>0$,
\[
|\lam(i;\zeta)|\leq C_2(1+i)
\]
for some constant $C_2$. Combining this with the result for $i\leq0$ proves \eqref{eq:lamest10}. Thus for each $i$ the series solution for $\lam$ converges absolutely and uniformly in $\zeta$ for $\zeta\in\cR^+$. Since the iterates $H_k$ are continuous in this region and analytic in its interior, $\lam$ also has this property. The results for $\up$ follow in a similar fashion.

\section{Acknowledgements}

The author would like to sincerely thank Nalini Joshi and Frank Nijhoff for their excellent support and the many helpful discussions. This research was funded by the Australian Research Council grant DP0985615.


\begin{thebibliography}{99}

\bibitem{akns:73}
M.J.~Ablowitz, D.J.~Kaup, A.C.~Newell and H.~Segur, Method for solving the sine-Gordon equation, {\em Phys. Rev. Lett} {\bf 30} (1973) 1262-1264

\bibitem{al:75}
M.J.~Ablowitz and J.~Ladik, Nonlinear differential-difference equations, {\em J. Math. Phys.} {\bf 16} (1975) 598-603

\bibitem{al:76}
M.J.~Ablowitz and J.~Ladik, Nonlinear differential-difference equations and Fourier analysis, {\em J. Math. Phys.} {\bf 17} (1976) 1011-1018

\bibitem{al:77}
M.J.~Ablowitz and J.~Ladik, On the solutions of a class of nonlinear partial difference equations, {\em Stud. Appl. Math.} {\bf 57} (1977) 1-12

\bibitem{abs:03}
V.E.~Adler, A.I.~Bobenko and Yu.B.~Suris, Classification of Integrable Equations on Quad-Graphs. The Consistency Approach, {\em Commun. Math. Phys.} {\bf 233} (2003) 513Ð543

\bibitem{av:04}
V.E.~Adler and A.P.~Veselov, Cauchy Problem for Integrable Discrete Equations on Quad-Graphs, {\em Acta Appl. Math.} {\bf 84} (2004) 237Ð262

\bibitem{bs:02}
A.I.~Bobenko and Y.B.~Suris, Integrable systems on quad-graphs, {\em Int. Math. Res. Not. IMRN} {\bf 11} (2002) 573-611

\bibitem{bpps:01}
M.~Boiti, F.~Pempinelli, B.~Prinari and A.~Spire, An Integrable Discretization of KdV at Large Times, {\em Inv. Prob.} {\bf 17} (2001) 515Ð526

\bibitem{b:12}
S.T.J.~Butler, Multidimensional inverse scattering of integrable lattice equations, {\em Nonlinearity} {\bf 25} (2012) 1613-1634

\bibitem{bj:10}
S.T.J.~Butler and N.~Joshi, An Inverse Scattering Transform for the Lattice Potential KdV Equation, {\em Inv. Prob.} {\bf 26} (2010) 115012

\bibitem{c:73}
K.M.~Case, On Discrete Inverse Scattering Problems II*, {\em J. Math. Phys} {\bf 14} 7 (1973) 916-920

\bibitem{ck:73}
K.M.~Case and M.~Kac, A Discrete Version of the Inverse Scattering Problem, {\em J. Math. Phys} {\bf 14} 5 (1973) 594-603

\bibitem{dt:79}
P.~Deift and E.~Trubowitz, Inverse Scattering on the Line, {\em Comm. Pure and App. Math} {\bf  32} (1979) 121-251

\bibitem{f:74}
H.~Flaschka, On the Toda Lattice. II, {\em Progress of Theoretical Physics} {\bf  51} 3 (1974) 703-716

\bibitem{g:66}
F.D.~Gakhov, Boundary Value Problems, Pergamon Press (1966)

\bibitem{ggkm:67}
C.S.~Gardner, J.M.~Greene, M.D.~Kruskal and R.M.~Muira, Method for solving the Korteweg-de Vries equation, {\em Phys. Rev. Lett.} {\bf 19} (1967) 1095-1097

\bibitem{ggkm:74}
C.S.~Gardner, J.M.~Greene, M.D.~Kruskal and R.M.~Muira, Korteweg-de Vries Equation and Generalizations. VI. Methods for Exact Solution, {\em Comm. Pure Appl. Math.} {\bf 27} (1974) 97-133

\bibitem{hz:09}
J.~Hietarinta and D.~Zhang, Soliton solutions for ABS lattice equations: Casoratians and bilinearization, {\em J. Phys. A} {\bf 42} (2009) 404006

\bibitem{h:77}
R.~Hirota, Nonlinear Partial Difference Equations I-III, {\em J. Phys. Soc. Japan} {\bf  43} (1977) 1424-33 

\bibitem{kn:80}
I.~Krichever and S.~Novikov, Holomorphic bundles over algebraic curves and nonlinear equations, {\em Russ. Math. Surv.} {\bf 35} (1980) 53-79

\bibitem{lp:07}
D.~Levi and M.~Petrera, Continuous Symmetries of the Lattice Potential KdV Equation, {\em J. Phys. A: Math. Theor.} {\bf 40} (2007) 4141-4159

\bibitem{m:90}
R.E.~Mickens, Difference Equations: Theory and Applications (2nd Edition), Van Nostrand Reinhold (1990)

\bibitem{n:02}
F.W.~Nijhoff, Lax Pair for the Adler (Lattice KricheverÐNovikov) System, {\em Phys. Lett. A} {\bf 297} (1Ð2) (2002) 49Ð58

\bibitem{na:10}
F.W.~Nijhoff and J.~Atkinson, Elliptic $N$-Soliton Solutions of ABS Lattice Equations, {\em Int. Math. Res. Notices} {\bf 20} (2010) 3837-3895

\bibitem{nah:09}
F.W.~Nijhoff, J.~Atkinson and J.~Hietarinta, Soliton Solutions for ABS Lattice Equations: I. Cauchy Matrix Approach, {\em J. Phys. A: Math. Theor.} {\bf 42} (2009) 404005

\bibitem{nqc:83}
F.W.~Nijhoff, G.R.W.~Quispel and H.W.~Capel, Direct Linearization of Nonlinear Difference-Difference Equations, {\em Phys. Lett. A} {\bf 97} (1983) 125-8

\bibitem{nw:01} 
F.W.~Nijhoff and A.J.~Walker, The Discrete and Continuous PainlevŽ Hierarchy and the Garnier System, {\em Glasgow Math. J.} {\bf 43A} (2001) 109Ð123

\bibitem{pnc:90}
V.~Papageorgiou, F.W.~Nijhoff and H.~Capel, Integrable mappings and nonlinear integrable lattice equations, {\em Phys. Lett.} {\bf 147A} (1990) 106-114

\bibitem{qcpn:91}
G.R.W.~Quispel, H.W.~Capel, V.~Papageorgiou and F.W.~Nijhoff, Integrable mappings derived from soliton equations, {\em Physica A} {\bf 173} (1991) 243-266

\bibitem{r:02}
S.N.M.~Ruijsenaars, A New Class of Reflectionless Second-Order A$\Delta$Os and its Relation to Nonlocal Solitons, {\em Reg. Chaotic Dynamics} {\bf 7} 4 (2002) 351-391

\bibitem{s:99}
A.~Shabat, Third Version of the Dressing Method, {\em Theo. and Math.l Phys.} {\bf 121} 1 (1999) 1397-1408

\bibitem{s:02}
A.~Shabat, Discretization of the Schr\"odinger Spectral Problem, {\em Inverse Problems} {\bf 18} (2002) 1003-1011

\bibitem{zs:72}
V.E.~Zakharov and A.~Shabat, Exact theory of two-dimensional self-focusing and one-dimensional self-modulation of waves in nonlinear media, {\em Sov. Phys. JETP} {\bf 34} (1972) 62-69

\end{thebibliography}
\end{document}